\documentclass[journal,12pt,draftclsnofoot,onecolumn]{IEEEtran}
\usepackage{svg}

\usepackage{cite}

\usepackage{graphicx}
\ifCLASSINFOpdf
\else
\fi
\usepackage{amsmath}
\usepackage{algorithm}
\usepackage{algorithmic}
\ifCLASSOPTIONcompsoc
  \usepackage[caption=false,font=normalsize,labelfont=sf,textfont=sf]{subfig}
\else
  \usepackage[caption=false,font=footnotesize]{subfig}
\fi

\begin{document}
%
\title{Joint User Association and Power Allocation in Heterogeneous Ultra Dense Network via Semi-Supervised Representation Learning}

%

\author{Xiangyu Zhang,~\IEEEmembership{Student Member,~IEEE,}
        Zhengming Zhang,~\IEEEmembership{Student Member,~IEEE,}
        and~Luxi Yang~,~\IEEEmembership{Senior Member,~IEEE}}
\maketitle

\begin{abstract}
Heterogeneous Ultra-Dense Network (HUDN) is one of the vital networking architectures due to its ability to enable higher connectivity density and ultra-high data rates. However, efficiently managing the wireless resource of HUDNs to reduce the wireless interference faces challenges. In this paper, we tackle this challenge by jointly optimizing user association and power control. The joint user association and power control problem is a typical non-convex problem that is hard and time-consuming to solve by traditional optimization techniques. This paper proposes a novel idea for resolving this question: the optimal user association and Base Station (BS) transmit power can be represented by some network parameters of interest, such as the channel information, the precoding matrices, etc. Then, we solve this problem by transforming it into an optimal representation function learning problem. We model the HUDNs as a heterogeneous graph and train a Graph Neural Network (GNN) to approach this representation function by using semi-supervised learning (SSL), in which the loss function is composed of the unsupervised part that helps the GNN approach the optimal representation function and the supervised part that utilizes the previous experience to reduce useless exploration in the initial phase. Besides, we use the entropy regularization to guarantee the effectiveness of exploration in the configuration space. To embrace both the generalization of the learning algorithm and higher performance of HUDNs, we separate the learning process into two parts, the generalization-representation learning (GRL) part, and the specialization-representation learning (SRL) part. In the GRL part, the GNN learns a representation with a tremendous generalized ability to suit any scenario with different user distributions, which processes offline. Based on the learned GRL representation, the SRL finely turn the parameters of GNN on-line to further improving the performance for quasi-static user distribution. Simulation results demonstrate that the proposed GRL-based solution has higher computational efficiency than the traditional optimization algorithm. Besides, the results also show that the performance of SRL outperforms the GRL.

\end{abstract}
\begin{IEEEkeywords}
Resource allocation, heterogeneous ultra dense network, graph neural network, semi-supervised learning.
\end{IEEEkeywords}

%
\IEEEpeerreviewmaketitle

\renewcommand{\thefootnote}{}
\footnotetext{
This work was supported by the National Natural Science Foundation of China under Grants 61971128 and U1936201, and the National Key Research and Development Program of China under Grant 2020YFB1804901.

X. Zhang, Z. Zhang, and L. Yang are with the National Mobile Communications Research Laboratory, School of Information Science and Engineering, Southeast University, Nanjing 210096, China, and also with Purple Mountain Laboratories, Nanjing 211111, China (e-mail: xyzhang@seu.edu.cn; zmzhang@seu.edu.cn;  lxyang@seu.edu.cn).}

%

\section{Introduction}
\IEEEPARstart{W}{ith} the progress of Internet of Everything, the next-generation mobile communication systems are expected to provide higher data rates, higher connectivity density, and ultra-low latency communication. To meet these needs, extensive flexible deployment of different kinds of small base stations (BSs) to enable more connection, namely, HUDNs, has been envisioned as one of the key technologies in 5G and 6G \cite{andrews_what_2014}\cite{saad_vision_2019}. Generally, HUDNs is an evolving architecture that comes from the legacy of dense networks \cite{andrews_are_2016}. Nevertheless, the mmWave\cite{xiao_millimeter_2017}, the massive MIMO\cite{larsson_massive_2014} and some new technologies bring the HUDNs with some new characteristics, such as cell-less, user-centric, and luxurious types of BS \cite{kamel_ultra-dense_2016}. These new characteristics enable HUDNs to have more flexibility in scheduling the wireless radio resource in the time domain, frequency domain, spatial domain. With efficiency HUDNs management strategies, the HUDNs would be to embrace higher spectral efficiency and further improve the HUDNs capacity. The prior works \cite{zheng_optimal_2017, qin_user_2018, xiao_joint_2018, kurras_spatial_2015,liao_model-driven_2020,cao_resource_2020,caidan_zhao_coloring-based_2016} demonstrated that optimizing the resources, such as power\cite{zheng_optimal_2017}, spectrum \cite{qin_user_2018}, and time \cite{xiao_joint_2018}, could significantly suppress the interference and improve performances in terms of  throughput capacity, energy efficiency, load balancing, etc. 

Among manageable wireless communication resource dimensions, the user association and power control play a pivotal role in HUDNs\cite{teng_resource_2019}. On the one hand, the interference, including inter-tier and cross-tier, is directly calculated by the user association matrices and BSs' transmit power. On the other hand, the user association and power control strategies are the most straightforward way to solve the load imbalanced problem caused by uneven UE and BS distribution \cite{zhou_joint_2016}. However, the joint user association and power control problem face the following three challenges:
\begin{itemize}
\item First of all, this problem contains two sub-problems: user association problem and power control problem. Both of them are highly non-convex problems that are hard to convert into convex problems and be solved by using the traditional optimization-based method. 

\item Besides, these two sub-problems are highly coupled with each other. The transmit power of BS impacts the received signal strength of UE and primarily determines the feasible user association set. The user association determines the resources that BS allocates to each UE. For these reasons, this problem can not be easily decomposed and solved separately. Besides, the traditional alternating optimization method, which requires a large number of iterations for reaching convergence.

\item Additionally, this problem is a large-scale optimization problem since the number of UEs and BSs in HUDNs is generally large. Meanwhile, due to the BS's heterogeneity, the feasible region for transmitting power of each BS could be different, which makes the feasible region complicated to explore. 
\end{itemize}

To overcome these limitations, we propose a novel approach for addressing joint user association and power allocation problem. We assume that a function can represent the relationship between corresponding optimal resource allocation and the network parameters. Under this assumption, we can solve this problem by formulating the optimal representation function. The main novelty of this paper is that we utilize the graph neural network as the representation approximating function with building the HUDNs as a heterogeneous graph model and combine the data-driven (supervised learning) and model-driven (unsupervised learning) training method to train the GNN for approximating the optimal representation and solving this problem. The proposed methodology has the advantage that the solving process is end-to-end, simplifying the problem, and algorithm design. This paper further considers improving the calculation efficiency of the algorithm through the combination of the off-line training (the GRL part) and on-line training (the SRL part). We show that our proposed method can significantly improve the data rate of all UEs.

\subsection{Related Works}
Plenty of prior works have addressed the problems of user association and power allocation. As for user association, Zhou \cite{wen_user_2017} utilized an ergodic scheme to find the optimal solution in UDN. However, the heavy computation limits the scope of application of this method. The paper \cite{siddique_channel-access-aware_2016} developed a tractable mathematical framework of Heterogeneous Network. They proposed a user association scheme by jointly considering the traffic load and channel quality through analysis of the performance. Power allocation of UDN had been considered in \cite{zheng_optimal_2017}, in which a novel dynamic-pricing game is utilized to maximize the sum-rate. 

Beyond that, the multi-dimensional coordination resource allocation of HUDNs has been given much more attention. The paper \cite{shen_distributed_2014} jointly considered the user association problem with power control and beamforming in the Heterogeneous Networks and solved the problem with the distributed pricing update strategy. In \cite{zhang_energy_2017}, user association and power allocation in mm-wave-based UDNs were considered as a mixed-integer programming problem. Through relaxing the integer  variable into continuous variables, the problem was solved by Lagrangian Dual Decomposition. A joint power allocation and user association strategy in HUDNs using non-cooperative game theory was developed in \cite{khodmi_joint_2019}. The proposed game was divided into two sub-games, the Backhaul Game and the Access Game. The Backhaul Game was implemented between BS and relay nodes (RNs) in the backhaul links, and the Access Game was implemented between the BS/RNs and UEs in the access links.

Except for the mentioned works focusing on the optimization-based method, e.g., the convex theory, the graph theory, and the game theory, the growing enthusiasm of reinforcement leaning develops a brand new way to solve these problems. Some the reinforcement-leaning-based method had been proposed in \cite{zhou_deep-learning-based_2018,li_user_2019,cheng_learning-based_2020,ding_deep_2020}. Modeling the user association and power allocation problem as a Markov Decision Problem, the reinforcement-leaning-based method find the desired  solution by repeatedly interacting with the environment, with less environment information requirements.


It is worth emphasizing that both optimization-based methods and reinforcement-leaning-based methods have their limitations \cite{teng_resource_2019}. For the optimization-based method, solving resource allocation problems always need precise and analytically tractable models, such as the channel model, antenna model, and local environment model, which is hard to obtain in a real environment. Besides, the closed-form expression between the manageable resource and the optimization criteria is hard to derive. As for the reinforcement-leaning-based method, the sample inefficient could probably limit the practical application. Beyond that, the solution of these algorithms could be hard to get for high time complexity, which impedes these algorithms implemented in the real environment.  

\subsection{Contribution}
This paper considers a typical two-tier HUDNs network, where the macro BS and small BS cooperatively service UE in the coverage area of the urban environment using the same frequency band. Our objective is to train a representation function between statistics channel information and the configuration of user association and transmitting power of BS to maximize the sum of effective data rates, which is defined in section II. In our paper, we quote the idea of the radio map \cite{bi_engineering_2019}, which is the geographical statistics channel information as a function of location. We assume, when the BSs' location and the environment are unchanged, the statistics channel information only relates to the UEs' distribution and can be extracted from the radio map. This assumption suit for the most scenario. We define a circumstance as an area where the BSs' location and the environment are quasi-static and an event as we get the stable statistics channel information to generate the configuration between the BS reconfiguration period. The main contributions of this paper are as follows:

\begin{itemize}
\item We formulate the resource allocation problem as an optimization problem and transform it into a representation function formulating problem. We utilize a learning method to found the optimal representation function, in which we train a neural network through a gradient back-propagation algorithm to approximate the optimal representation function iteratively. The result of our algorithm proves that the neural network has learned effective representation and shows excellent performance than the traditional method.

\item To utilize the structure information implicit in HUDNs, we creatively formulate a brand new heterogeneous graph architecture to represent the relationship between UE and BS in HUDNs. Our formulated graph architecture embraces the feature that the one-order neighbor and second-order neighbor are the same types of nodes. Based on that, we employ the GNN mentioned in the paper \cite{hamilton_inductive_2018} as the feature extractor of our neural network since it has a remarkable ability to process the information on the graph. However, the GNN can not be directly utilized in the heterogeneous graph. We extend the algorithm by utilizing two different aggregators to gather different order neighbors' information, ensuring the aggregate operation can be applied in a heterogeneous graph and perfectly match the HUDNs model we modeled before.

\item For maximum the sum of effective data rate, we use the objective as the loss to train the network, which is unsupervised learning. However, training only with unsupervised learning is ineffective. We design a special semi-supervised learning algorithm to train our proposed neural network, where the loss function is the dynamical weighted sum consisting of three parts, the supervised learning part, the unsupervised learning part, and the exploration part. The supervised part can efficiently reduce the useless exploration in the initial training part by using some labels that generate by some existing algorithms. The exploration part is to maximize the entropy of user association output to guarantee the effectiveness of exploration in the configuration space, which is a widely accepted fact in reinforcement learning \cite{schulman_proximal_2017}.

\item To pursue effective data rates and calculation efficiency, we decompose the learning process into the GRL and SRL parts. In the GRL part, we train the neural network off-line to learn the generalized representation for an environment suitable for any event in this circumstance. The GRL part only needs to be trained once for a circumstance, which is calculation efficient when using on-line. The SRL finely tune neural network parameter to let the neural network fit into an event, which could further improve the total sum date rate.  

\item Numerical results are provided. We present the convergence performance of our method to show the effectiveness of our algorithm. Also, we compare the total effective date rate of our algorithm with some optimization-based approaches. The cumulative distribution functions of date rate for GRL and SRL in different UE density are also presented to verify the result of the SRL part. Finally, the time complexity of different approaches is provided.

\end{itemize}
\subsection{Organization of This Paper}
The rest of the paper is structured as follows. Section II  introduces the system model, the channel model, and the formulation of the optimization problem. In section III, we briefly introduce the background of semi-supervised learning, our modeling of graph, and graph neural network. Section IV establishes a graph model for heterogeneous UDN and uses a semi-supervised graph neural network to solve the joint optimization problem. Section V analyzes the performance of our proposed algorithm through simulation. Finally, we conclude that our methods and outlook the future research direction at the end of paper.

\section{System Model And Problem Formulation}
In this section, we present the basic information of scenario, communication model and formulate the joint user association and power control problem as an optimization problem.
\subsection{Scenario}
\begin{figure}[h]
\centering
\includegraphics[scale=0.24]{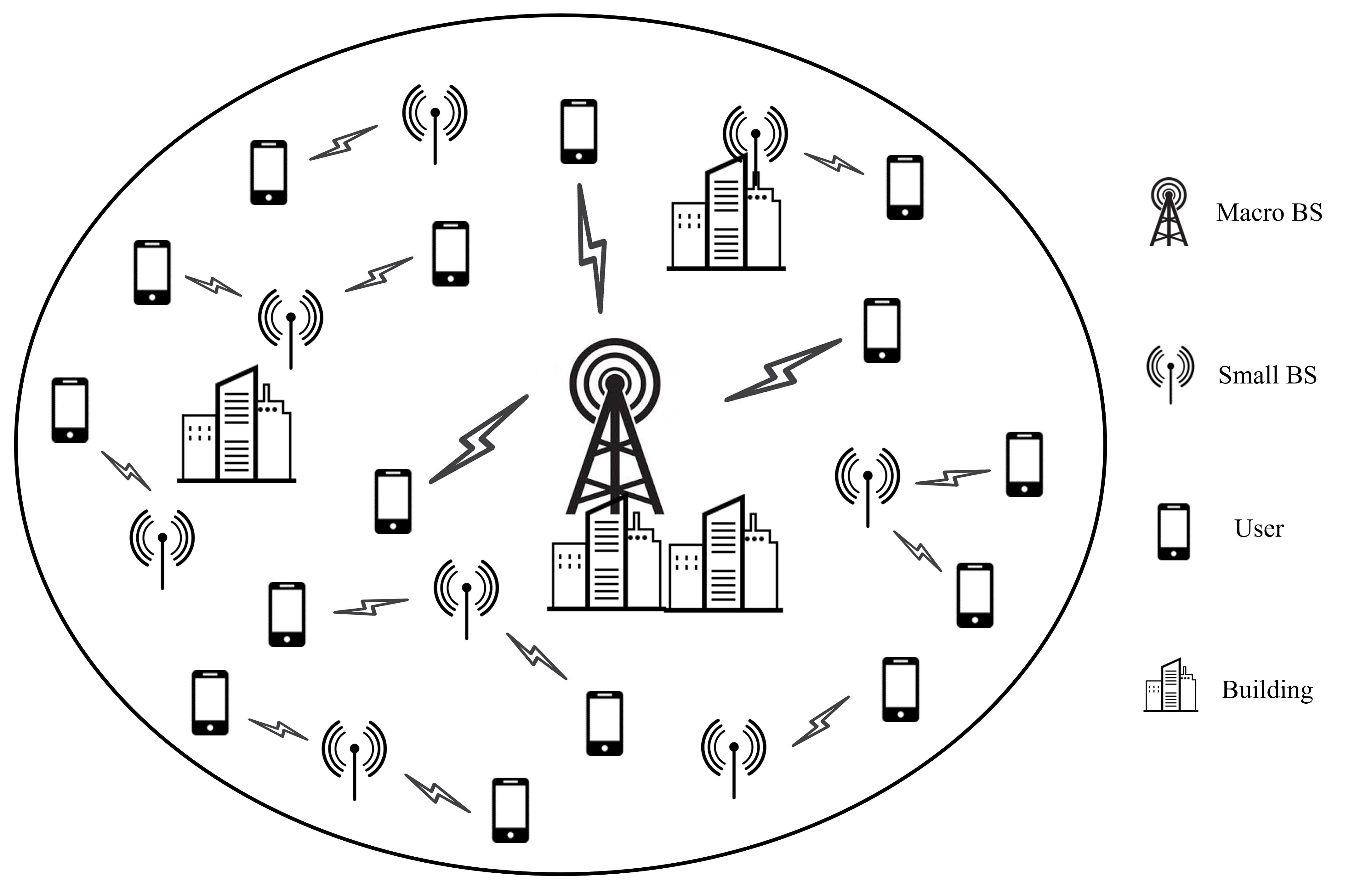}
\caption{Architecture of a heterogeneous Ultra-Dense Network.}
\end{figure}

As shown in Fig.1, we investigate the joint user association and power allocation problem in the downlink of the HUDNs scenario, where ultra-dense small BSs and Marco BSs cooperatively service the UE in the converging area. We consider a realistic city environment where the buildings impact the propagation of the signal. The small BSs randomly distribute in the scenario, and the Marco BSs are evenly deployed. Without a specific claim, the BSs consist of small BSs and Marco BSs, which are represented by set $\mathcal{J}=\{1, \ldots, J\}$. In the $\mathcal{J}$, the first $M$ items denote the Marco BSs, and the last of $N$ items are the small BS. We assume that the small BSs and macro BSs interfere with each other for sharing the same frequency band, and the UEs connected to the same BS can avoid interference. We denote the single-antenna UEs as $\mathcal{I}=\{1, \ldots, I\}$. 

In practice, the adjustment of use association and down-link power occur in two time spans: (i) the instantaneous adjustment, which adjusts within each time slot and (ii) the long-term adjustment, which adjusts the configuration when the users move or the user randomly access or leave the network. In the long-term reconfiguration, the network changes the configuration over several time slots or time frames. We focus on the scenario of the long-term reconfiguration and assume the users remains unchanged between the BS reconfiguration period.


%
%
%


We use the binary integer variable $x_{i, j}$ to indicate the association variable between UE $i$ and BS $j$. If UE $i$ is associated with BS $j$, $x_{i, j}=1$, otherwise $x_{i, j}=0$. Let $p_{j}$ be the transmit power from BS $j$ and $g_{i, j}$ be the power gain from BS $j$ to UE $i$. The SINR of UE $i$ receiving from BS $j$ can be written as
\begin{equation}
\operatorname{SINR}_{i, j}=\frac{p_{j} g_{i, j}}{\sum_{n \in \mathcal{J} \backslash\left\{j\right\}} p_{n} g_{i, n}+\sigma^{2}},
\end{equation}
where $\sigma^{2}$ is the variance of Additive White Gaussian Noise (AWGN). 
Then the effective data rate for UE $i$ from BS $j$ can be calculated by
\begin{equation}
\gamma_{i, j}=\frac{B}{K_{j}} \log _{2}\left(1+\operatorname{SINR}_{i}\right),
\end{equation}
where $B$ is the system bandwidth, $K_{j}=\sum_{k \in \mathcal{I}} x_{i, j}$ is the total number of users associated with BS $j$, and thus each user can receive $1 / K_{j}$ of the total frequency band available.

\subsection{Channel Model}

Traditionally, the power gain $g_{i, j}$ can be written as 
\begin{equation}
g_{i, j} = \beta_{m}(w) G_{m}(w) \tilde{h}_{m}(t),
\end{equation}
where $\beta_{m}(w)$ denote the large-scale channel gain, which generally depends on the distance $w$ between UE and BS. And $G_{m}(w)$ and $\tilde{h}_{m}(t)$ are the BS antenna gain and the small-scale fading, respectively. Let $d$ denotes two-dimensional (2D) distance and $h$ for representing the absolute antenna height difference between the BS and the UE, the three-dimensional (3D) distance between the BS and the UE can be expressed as $w=\sqrt{d^{2}+h^{2}}$. 

For the long-term adjustment, the network often rely on the statistics channel information rather than the instantaneous channel information to provide configuration that comprehensively consider next period. Thus, we quote the idea of the radio map \cite{bi_engineering_2019}, which is the geographical statistics channel information as a function of location. As mentioned before, we assume the BS can extract the statistics channel information of UE from the radio map according to the location of UE. In practice, the radio map can be measured and updated by instrument or estimated by a channel model \cite{lee_voronoi_2012} \cite{fu_factor_2015}. Nevertheless, for illuminating the effectiveness of our method, in our simulation, we use the practical two-stage 3GPP model in the period work\cite{ding_performance_2017} to generate the radio map, which will be discussed in Section V.

\subsection{Problem Formulation}
Before modeling the problem, we give the following constraints:

(1) User scheduling constraint: A user can only be associated with one BS at a time, therefor
\begin{equation}
\sum_{j \in \mathcal{N}} x_{i,j}=1, \quad \forall i \in \mathcal{K}_s.
\end{equation}


(2) Maximum power constraint: The maximum transmit power of each BS is $p_{max}$, therefore,
\begin{equation}
p_{j}\leq p_{max}, \quad \forall j \in \mathcal{N}.
\end{equation}

Our objective is to maximize the total effective rate over the coverage area through the joint optimization of power allocation and user association. The optimization objective function is 
\begin{equation}
\begin{aligned}
\mathop{\arg\max}_{\mathbf{x},\mathbf{p}}  & \, R =\sum_{i \in \mathcal{K}_s}\sum_{j \in \mathcal{N}} x_{i, j} \gamma_{i, j} \\
\text{s.t.}\ &\text{C1}: \sum_{j \in \mathcal{N}} x_{i, j}=1, \quad \forall i \in \mathcal{K} \\
&\text{C2}:  p_{j}\leq p_{max}, \quad \forall j \in \mathcal{N},
\end{aligned}
\end{equation}
where the $\mathbf{x}= [x_{i, j}]$ is the association matrix and $\mathbf{p}=[p_{j}]$ is the power vector.

In HUDNs, the problem (6) is a typical large-scale optimization problem, which contains both the continuous and disperse variables. Traditionally, the optimization-based method decomposes the problem (6) into two sub-optimization problems and iteratively solves these two problems. The solution of these methods could be hard to coverage into optimization, and the solving process is time-costly. These problems can be avoided in our learning method because the two sub-problems share one neural network, and the loss function for train the neural network is jointly impacted by $\mathbf{x}$ and $\mathbf{p}$. We will introduce the learning algorithm in section III and detail our method in section IV. 


\section{An Overview Of Semi-Supervised learning and Graph Neural Network}

This section aims to present some pivotal notations about Semi-Supervised learning and the Graph Neural Network that is to be used in the sequel of this paper. For a more comprehensive description, the readers are referred to the books\cite{bishop_pattern_2006} and \cite{liu_introduction_2020}.

\subsection{Semi-Supervised Learning}
The semi-supervised learning is the learning method between supervised learning and unsupervised learning. In practice, most of the semi-supervised learning has two views. The first view is to treat the semi-supervised leaning as unsupervised learning with the constraint from the supervised information. In this view, the semi-supervised learning is mainly focused on the clustering problem, which typically either modifies the objective clustering function or to learn the clustering measurement. Another view is seen as semi-supervised learning as an extension of supervised learning. Most research in this view focus on the classification problem that trains the classifier on both labeled data and unlabeled data, which can be formulated as:

\begin{equation}
\begin{aligned}
\log p(\mathcal{D} | \theta) &=\log \left(\prod_{i=1}^{l} p\left(\mathbf{x}_{i}, y_{i} | \theta\right) \prod_{i=l+1}^{l+u} p\left(\mathbf{x}_{i} | \theta\right)\right) \\
&=\sum_{i=1}^{l} \log p\left(y_{i} | \theta\right) p\left(\mathbf{x}_{i} | y_{i}, \theta\right)+\sum_{i=l+1}^{l+u} \log p\left(\mathbf{x}_{i} | \theta\right),
\end{aligned}
\end{equation}
Where $
\mathcal{D} = \left\{\left(\mathbf{x}_{1}, y_{1}\right), \ldots,\left(\mathbf{x}_{l}, y_{l}\right), \mathbf{x}_{l+1}, \dots, \mathbf{x}_{l+u}\right\}
$ is the data set contain labeled data and unlabeled data. The first term is the supervised learning for labeled data, and the second term for unlabeled data.

\subsection{Graph Model and Graph Neural Network}
The graph is a ubiquitous non-Euclidean data structure that extensively exist in pharmacy, chemistry and related field. Normally, a graph $G = (V,E)$ comprises a set of nodes $V$ and a set of edges $E$ that connects two nodes and describes the relationship between different nodes. Because graphs have strong expressive power and can be used as an extension of many systems, they have been applied to many research fields such as social sciences (social networks), natural sciences (physical systems and protein interaction networks), and knowledge graphs. As a unique non-Euclidean data structure for machine learning, graph analysis mainly focuses on node classification, link prediction, and clustering. Traditionally, from the view of nodes' kinds, graph can be broadly classified in homogeneous graph and heterogeneous graph. Homogeneous graph means containing only one type of node and relationship in the graph, which is the most simplified case of actual graph data. The information on this type of graph data is all contained in the adjacent matrix. In contrast, heterogeneous graphs refer to more than one type of node or relationship in the graph. 



To extend the deep learning from Euclidean data to non-Euclidean data, Graph neural network (GNN) has emerged in amount of research work\cite{wu_comprehensive_2019}. Showing the convincing performance and high interpretability, GNN has become a widely used graph theory analysis method in recent years. GNN is a connection model that captures graph dependencies through message passing between nodes of the graph. Unlike standard neural networks, graph neural networks retain a state that can represent information of arbitrary depth from their neighborhood. 

The basic motivation of GNNs is the convolutional neural network\cite{lawrence_face_1997}. CNN can extract multi-scale local spatial features and combine them to construct a highly expressed representation. However, CNN can only handle conventional Euclidean data, such as images (2D grids) and text (1D sequences), and these data structures can be regarded as particular cases of graphs. Another motivation comes from graph embedding, which learns to represent graph nodes, edges, or subgraphs in low-dimensional vectors. In the field of graph analysis, traditional machine learning methods usually rely on hand-designed features and are limited by their flexibility and high cost. Based on CNN and graph embedding, GNN is proposed to model the information in the graph structure.

GNN originated from the paper\cite{scarselli_graph_2009}, which is based on the fixed point theory, namely Banach's Fixed Point Theorem. The earliest GNN mainly dealt with graph theory problems in a strict sense such as molecular structure classification. But in fact, data with European structure, such as images or text, and many common scenes can also be converted into graph representations, and then graph neural network technology can be used for modeling. In 2013, based on Graph Signal Processing, Bruna firstly proposed the convolutional neural network based on spectral-domain and spatial-domain on the graph in the literature\cite{bruna_spectral_2014}, which draw many scholars putting attention on graph convolution methods based on spatial-domain. The Graph Convolutional Network (GCN) is the first spectral-domain GNN that proposed by paper \cite{kipf_semi-supervised_2016}. The basic idea of GCN is that learning on graph-structured data via an efficient variant of convolutional neural networks, which is a localized first-order approximation of spectral graph convolutions. 

The GCN is a typical transductive learning algorithm \cite{joachims_transductive_nodate} that need all nodes to participate in training to learn the the global information of the graph structure. However, it is hard to train on large graph. Contrastively, the inductive learning framework shows advantage in large graph by learning more general embedding for all nodes. The most representative work is the GraphSAGE proposed by Hamilton\cite{hamilton_inductive_2018}. GraphSAGE is the abbreviation of Graph SAmple and aggreGatE. It is to learn a node representation method, that is, how to sample and aggregate vertex features from a local neighbor of a vertex, rather than training a separate embedding for each vertex. The algorithm flow can be divided into three steps: (1) sample the neighboring vertices of each vertex in the graph; (2) aggregate the information contained in the neighboring vertices according to the aggregation function; (3) obtain the vector representation of each vertex in the graph for downstream tasks use.

The GNN for heterogeneous graph was mentioned in \cite{wang_heterogeneous_2019}. In this paper, a new Heterogeneous Graph Attention Network (HAN) based on the attention mechanism was proposed, which can be widely used in heterogeneous graph analysis. The embedding of nodes in heterogeneous graphs mainly focuses on the structural information based on meta-path, which is a sequence from one node to another. The HAN model follows a hierarchical attention structure, from node-level attention to semantic-level attention. Node-level attention learning is based on the importance of meta-path's node neighbors, and semantic-level attention learns the importance of meta-path and merges semantic information. Because of the layered attention mechanism added to HAN, it has a better ability to explain heterogeneous graphs.




\section{Propose Method}

In this section, we present the details of our proposed algorithm for solving the problem in (7). Our proposed method is ordinarily established on the radio map containing plenty of channel information from every UE to BS. Based on the radio map, the interference relationship between UE/ BS, UE/UE can be straightforwardly expressed. Hence, the HUDNs can easily be modeled as the graph model. Then, we formulate the problem (7) as a graph representation learning problem, whose objective is to learn a representation function that transforms the channel information of each UE and each BS into user association matrix $\bf{x}$ and power control vector $\bf{p}$. For the reason that representation functions are hard to design by hand, we employ GNN to approach these functions and train them with semi-supervised learning.


\subsection{Graph Model and Heterogeneous GNN for HUDNs}
\begin{figure}[h]
\centering
\includegraphics[width=5in]{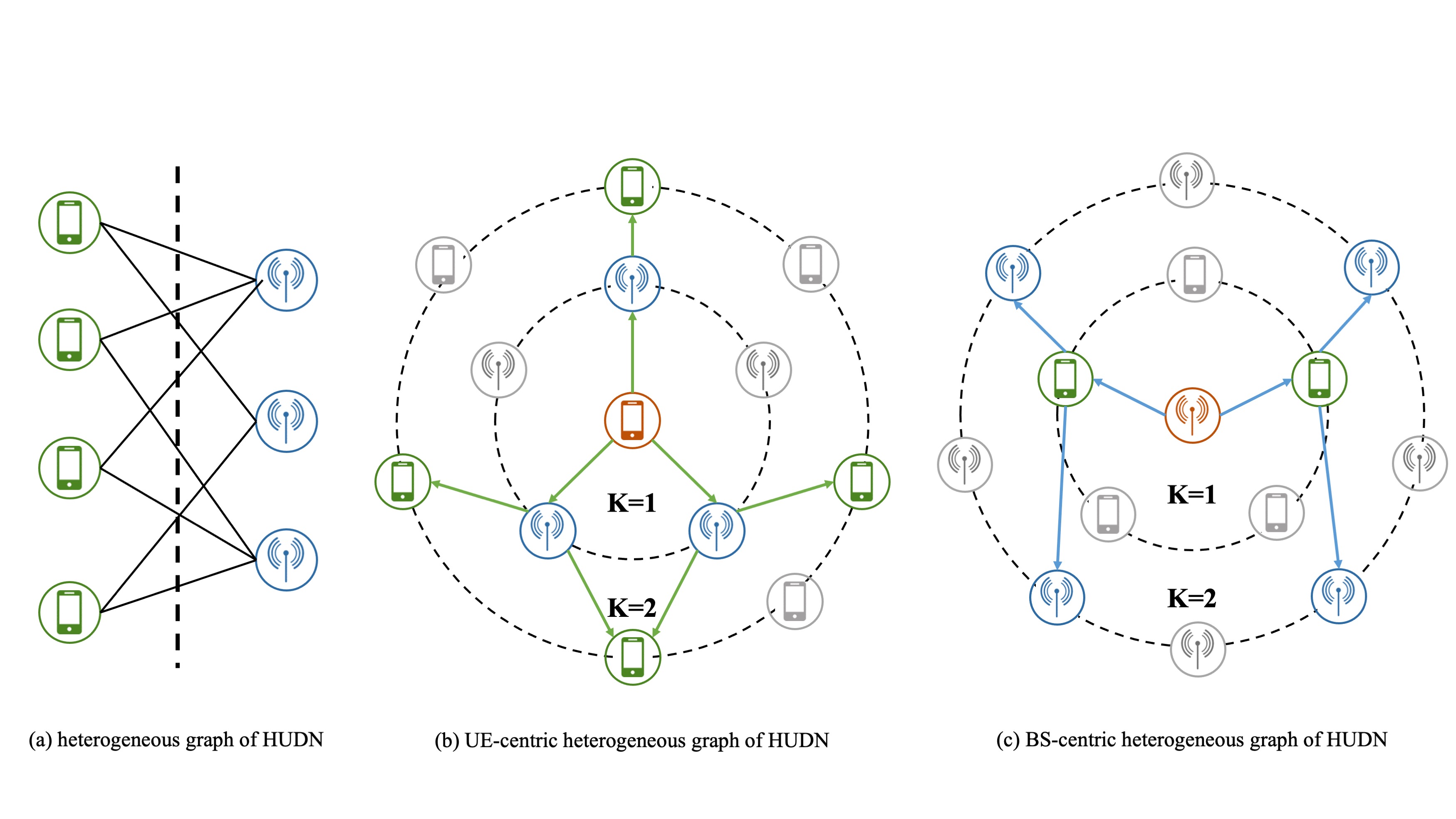}
\caption{Heterogeneous Graph and Heterogeneous Graph neural network of HUDNs.}
\end{figure}
The first step in our method is to build the HUDNs as a heterogeneous graph. As shown in Fig.2(a), we represent the UEs and BSs by two different kinds of nodes that contain different kinds of features. The features for both UE nodes and BS nodes are the statistics channel information extracting from the radio map but with different views. The features of UE nodes are the channel to every BS. However, The BS nodes' features are the channel to every UE. The edge of the graph only exists between the UE nodes and the BS nodes when the UE can be detected. For more straightforward expression, we utilize the definition of one order neighborhood and two order neighborhood to define the connection relationship between UE and BS. The one order neighborhood of UE is the BS that has edge connect, shown in Fig.2(b) for $K=1$, the second-order neighborhood of UE is the UE that connects to the one order neighborhood, showed in Fig.2(b) for $K=2$. The corresponding neighborhood of BS is the same, shown in Fig.2(c).

\begin{figure}[h]
\centering
\includegraphics[width=5in]{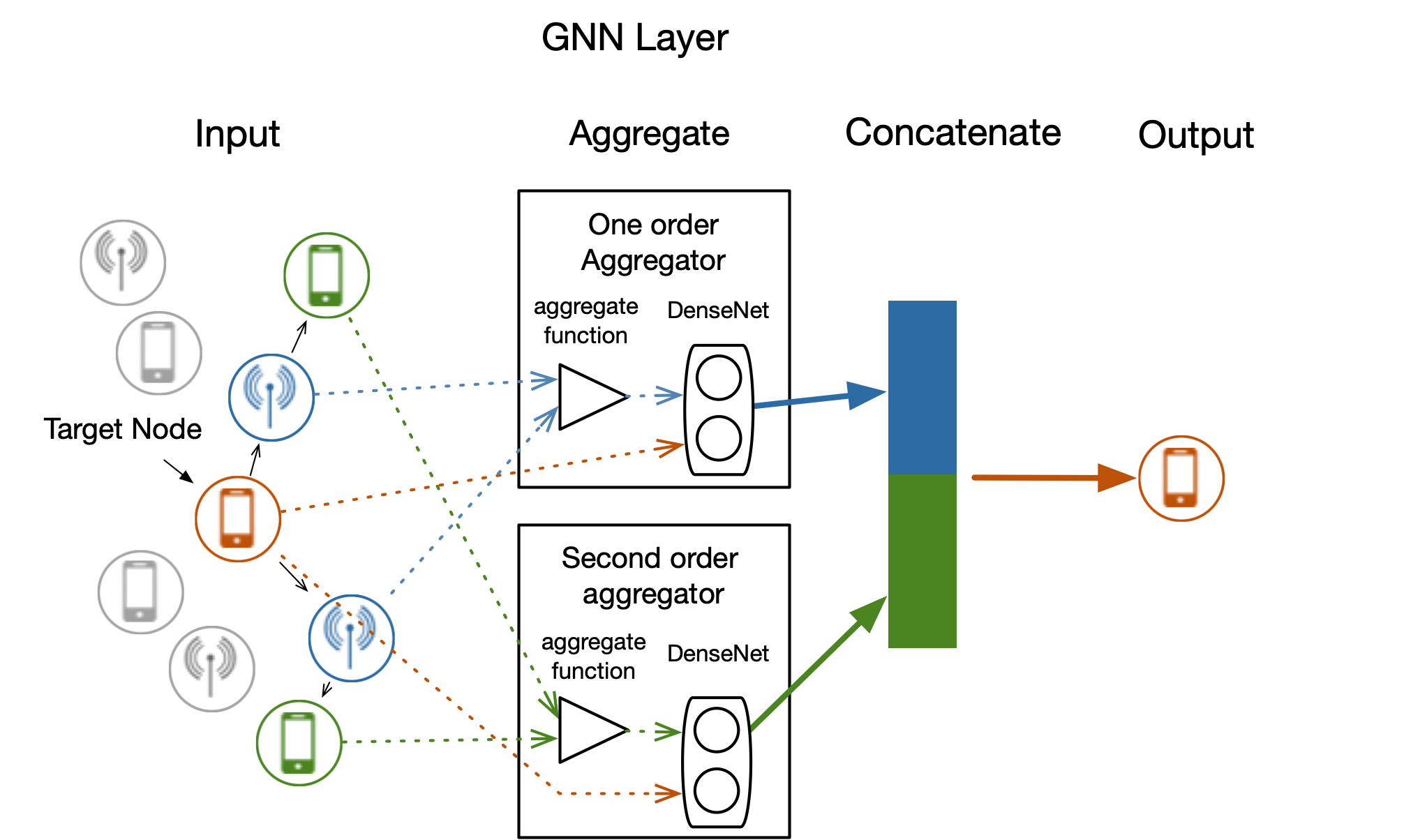}
\caption{The Heterogeneous GNN for HUDNs.}
\end{figure}

The key idea of GNN, such as GraphSAGE and GAT, extend CNN on the graph that trains several kernels to aggregate the neighbors' information and learn the representations. It is hard to fit the GraphSAGE into the graph of HUDNs directly. However, our formulated graph embraces the character that the one order neighbor and the second-order neighbor are the same types of nodes. Utilizing this characteristic, we extend the GraphSAGE to the heterogeneous graph by employing two different order aggregation kernels, namely: one-order and second-order aggregator, to extract the features from one-order and second-order neighborhood, respectively. And we call it Heterogeneous GraphSAGE (HGSAGE). As shown in Fig.3, in each layer, the aggregator sample corresponding order neighbors of target nodes and collect these nodes' features and aggregate them with an aggregate function. Then, we combine the aggregated features with the target node's feature and send it in to a fully connected neural network, to generate the embedding feature. The output of the HGNN layer is the combination of both order features. The summarization is shown in Algorithm 1. In the algorithm 1, the $\operatorname{CONCAT}(\mathbf{x,y})$ denotes the concatenation of two vector $\mathbf{x,y}$. 

\begin{algorithm}
\caption{HGSAGE}
\label{alg1}
\begin{algorithmic}[1]
\STATE \textbf{Input}: Heterogeneous Graph $G$; input feature ${\bf{x}}_v$; layer num $K$; one order neighborhood set of node $v_1 \in \mathcal{N}_1(v)$; second order neighborhood set of node $v_2 \in   \mathcal{N}_2(v)$; aggregate function $\mathcal{F}$; activation function $\sigma$. 
\STATE Randomly initialize the weight of one order aggregator and second order aggregator $\bf{W}_1$ $\bf{W}_2$ and sample a set of nodes $\bf{v}$ from the graph; 
\STATE ${\bf{h}}_v^0 \leftarrow {\bf{x}}_v$
\STATE \textbf{For} $k$ = 1 to $K$ \textbf{do}:
\STATE \quad \textbf{For} v = 1 $\in V$ \textbf{do}:
\STATE \qquad $
\mathbf{o}_{\mathcal{N}_1(v)}^{k} \leftarrow \mathcal{F}_{k}\left(\left\{\mathbf{h}_{u}^{k-1}, \forall u \in \mathcal{N}_1(v)\right\}\right)
$
\STATE \qquad $\mathbf{o}_{v}^{k} \leftarrow \sigma\left(\mathbf{W}_1^{k} \cdot \operatorname{CONCAT}\left(\mathbf{h}_{v}^{k-1}, \mathbf{o}_{\mathcal{N}_1(v)}^{k}\right)\right)$

\STATE \qquad $
\mathbf{s}_{\mathcal{N}_2(v)}^{k} \leftarrow \mathcal{F}_{k}\left(\left\{\mathbf{h}_{u}^{k-1}, \forall u \in \mathcal{N}_2(v)\right\}\right)
$
\STATE \qquad $\mathbf{s}_{v}^{k} \leftarrow \sigma\left(\mathbf{W}_2^{k} \cdot \operatorname{CONCAT}\left(\mathbf{h}_{v}^{k-1}, \mathbf{s}_{\mathcal{N}_2(v)}^{k}\right)\right)$

\STATE \qquad $\mathbf{h}_{v}^{k}\leftarrow\operatorname{CONCAT}(\mathbf{o}_{v}^{k},\mathbf{s}_{v}^{k} )$
\STATE \quad \textbf{End}

\STATE \quad $\mathbf{h}_{v}^{k} \leftarrow \mathbf{h}_{v}^{k} /\left\|\mathbf{h}_{v}^{k}\right\|_{2}, \forall v \in \mathcal{V}$
\STATE \textbf{End}
\end{algorithmic}
\end{algorithm}

\subsection{Semi-Supervised Learning for Joint User Association and Power Control in HUDNs}
In this subsection, we present our neural network that approximates the representation functions and the semi-supervised learning method. As mentioned before, the user association and power control are two couple problems. Meanwhile, these two problems extract the feature from the same radio map. For such reason, we employ one input and two output neural network architecture to jointly approximate the representation function. As shown in Fig.4, our network leverages two HGNN layers as the feature extractor. Then, two fully connected output layers are utilized to generate the user association and power control configuration. One with sigmoid output function generates the normalized power ${\bf{p}}$ of BSs. Another is to generate the user association configuration. To constrain $x_{n,k}$ as binary integer variable, satisfy the constraint (5) and keep the $x_{n,k}$ differentiable, we employ softmax output function with the low temperature parameter \cite{gao_properties_nodate}, which can be shown as:
\begin{equation}
x_{n,k}=\frac{\exp \left(z_{n,k} / T\right)}{\sum_{j=1}^{N} \exp \left(z_{j,k} / T\right)} \quad \forall n \in \mathcal{N},\forall k \in \mathcal{K}_s,
\end{equation}
where the $z_{n,k}$ is the output of User Association layer.

\begin{figure}[h]
\centering
\includegraphics[width=5in]{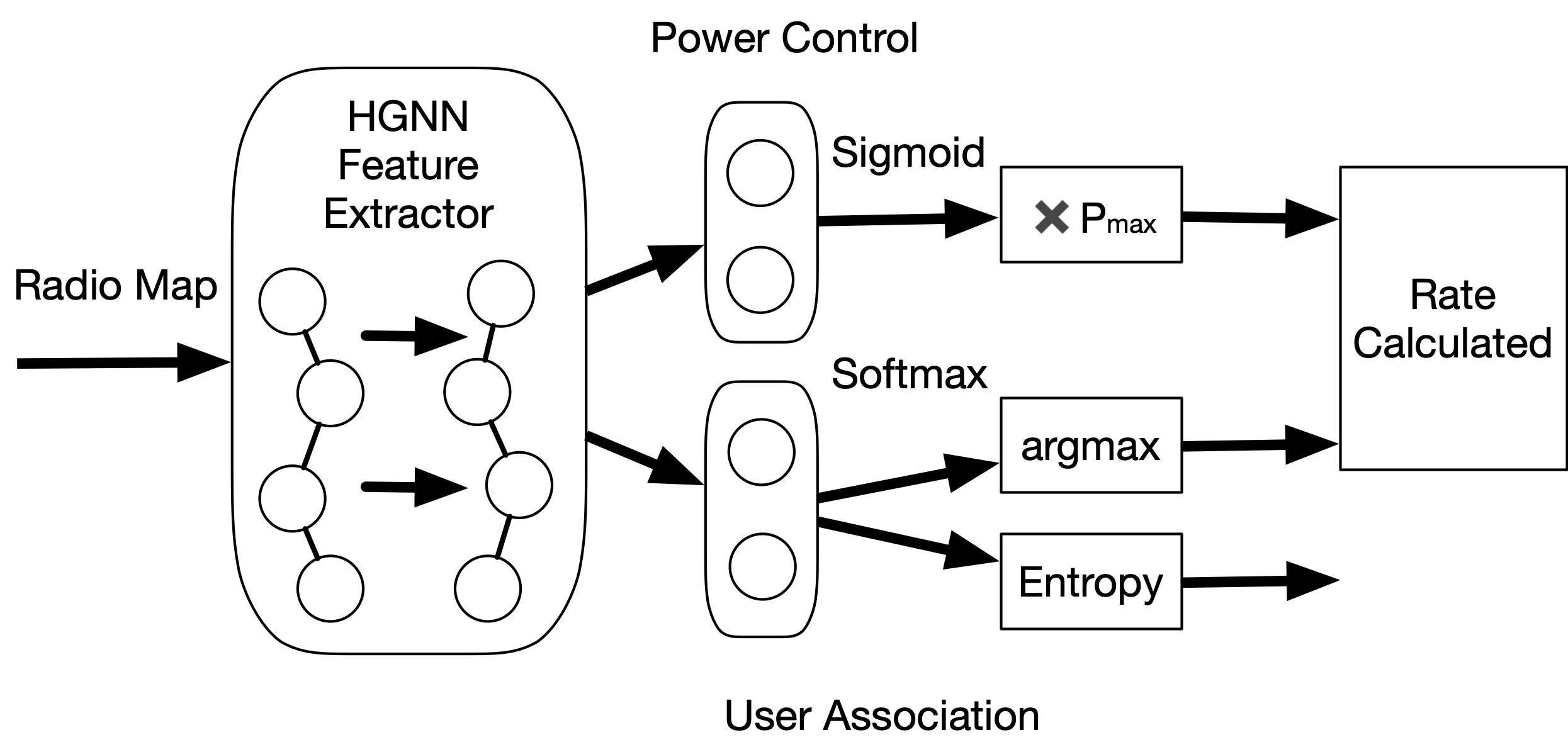}
\caption{The HGNN Architecture of Representation Functions.}
\end{figure}

Typically, the proposed representation learning problem is an unsupervised learning problem. To maximize the objective in problem (6), the loss for training the neural network is deservedly the objective function, which is the effective rate under the configuration that the neural network generates, which is $L_{u}=R$.

\textbf{\emph{Theorem1:}} When the output of softmax function approaches the one-hot vector, the $\frac{\partial {L}_u}{\partial \omega}=0$, which blocks the gradient back propagation and leads to this architecture of neural network failed to train.


\textbf{\emph{Proof1:}} The gradient of loss to weight is 
\begin{equation}
\frac{\partial {L}_u}{\partial \omega}=\sum_{i \in \mathcal{K}_s}\sum_{j \in \mathcal{N}}\left( \frac{\partial {L}_u}{\partial \gamma_{i, j}}\frac{\partial \gamma_{i, j}}{\partial \omega}x_{i, j}+\frac{\partial {L}_u}{\partial x_{i, j}}\frac{\partial x_{i, j}}{\partial \omega}\gamma_{i, j} \right).
\end{equation}

Then we focus on the softmax part
\begin{equation}
\frac{\partial x_{i, j}}{\partial \omega} = \frac{\partial x_{i, j}}{\partial z_{n,j}}\frac{\partial z_{n,j}}{\partial \omega} \quad  \forall n \in \mathcal{N}.
\end{equation}

The $\frac{\partial x_{i, j}} {\partial z_{n,j}} $  is derivated from  \cite{hinton_distilling_2015}, which is given by:

\begin{equation}
\begin{aligned}
\frac{\partial x_{i,j}}{\partial z_{i,j}} &=\frac{1}{T}\frac{\exp \left(z_{i,j}/T\right) \cdot\left(\Sigma-\exp \left(z_{i,j}/T\right)\right)}{\Sigma^{2}}=\frac{1}{T}x_{i,j}\left(1-x_{i,j}\right) \\
\frac{\partial x_{i,j}}{\partial z_{i,n}} &=-\frac{1}{T}\frac{\exp \left(z_{i,j}/T\right) \cdot \exp \left(z_{i,n}/T\right)}{\Sigma^{2}}=-\frac{1}{T}x_{i,j} x_{i,n}.
\end{aligned}
\end{equation}

As the temperature parameter of softmax declines, the $x_{i,j}, j \in \mathcal{N}$ approach the one-hot vector. Thereby, the gradient $\frac{\partial x_{i,j}}{\partial z_{i,j}}$, $\frac{\partial x_{i,j}}{\partial z_{n,j}}$ would approach to zero. The training would be hard to proceed.

We use the following way to alleviate the influence that low-temperature parameter brings. Firstly, we eliminate the negative effects of softmax approaching the output of power. As the unsupervised loss is calculated by two outputs of the neural network, which is the right side of equation (10), we can separate the training process into two parts, training the neural network through $p_j$ and $x_{i,j}$. When training the neural network through $p_j$, the $x_{i,j}$ can be treated as parameters, thus it can be approached by argmax function sequentially avoiding this effect. Then the deviation of the softmax approach will not affect the output of power. Besides, we normalize the output $z_{n,k}$ and choose a suit temperature parameter to enable the $\frac{\partial x_{i,j}}{\partial z_{i,j}} = 1$ as far as possible. For all this, the gradient that backpropagated through the softmax function still unstable for the reason that only one neural cell can work in each sample. Fortunately, the UEs in each sample connect to different BS, thus we can update the network using batch samples to reduce variation.



The graph of HUDNs is definitely a large graph. Learning on this graph is easily puzzled by the dimensionality curse, especially for the representation of user association that transforms the variable from $I \times J$ dimension to $I \times J$ dimension. Thus, learning through the unsupervised learning method thought the backpropagation\cite{sanger_optimal_1989} is inefficient and could lead to poor performance due to either falling into local optimum or drifting convergence\cite{sontag_backpropagation_nodate} \cite{beyer1999nearest}. In this paper, we borrow the ideas from the semi-supervised clustering \cite{bai_semi-supervised_2020} that we treat the user association as a classification problem and provide some label information (user prior association) to guide the search process. In another word, we constrain some UEs' associate to a specific BS. And we use (12) to be the supervised loss function of our HGNN to learn the user association solutions. 
\begin{equation}
{L}_s = -\sum_{j \in {\mathcal{K}}_s}\sum_{i \in \mathcal{N}}   y_{ij} \ln x_{ij},
\end{equation}
where the $y_{i,j}$ is the label for $x_{i,j}$, which is the solution of optimization-based method. $y_{i,j}=1$ when the UE $j$ connected to BS $i$, otherwise, $y_{i,j}=0$.

Then the loss function for our proposed method comprises two items: the supervised part, the unsupervised part, which can be showed as 
\begin{equation}
{L}= w_s{L}_{s}-w_rR,
\end{equation}
where $w_s$ and $w_r$ are weight for the supervised part and the unsupervised part.

\subsection{Generalization-Representation Learning and Specialization-Representation Learning}
\begin{algorithm}
\caption{Generalization Representation Learning Algorithm}
\label{alg2}
\begin{algorithmic}[1]
\STATE \textbf{Input}: The radio map $\mathcal{M}$; Training times $T$; Batch size $B$; learning rate $\eta$; the decay factor of learning rate $ \tau$; The window length $l$.
\STATE Generate the Heterogeneous Graph $G$.
\STATE Create an empty memory set $M$ and the average best rate $r_b=0$.
\STATE Build the HGNN neural network and initialize the parameter of HGNN.
\STATE \textbf{For} t = 1 to $T$ \textbf{do}:
\STATE \quad ${L}_b = 0$.
\STATE \quad \textbf{For} b = 1 to $B$ \textbf{do}:
\STATE \qquad Activate $K$ UEs and extract corresponding radio map $\mathcal{M}_s$ from $\mathcal{M}$.
\STATE \qquad ${\bf{h}}_v^0 \leftarrow \mathcal{M}_s$.
\STATE \qquad Calculate the user association $\mathbf{x}$ and power control $\mathbf{p}$ configuration thought HGNN \\
\qquad neural network.
\STATE \qquad Calculate the loss function ${L}$ of (13).
\STATE \qquad ${L}_b \leftarrow {L}_b + {L}$.
\STATE \qquad Add the rate $R$ into set $M$.
\STATE \quad \textbf{End For}
\STATE \quad Calculate the average rate $r_n$ of recent $l$ elements in memory set $M$.
\STATE \quad \textbf{If} $r_n > r_b$
\STATE \qquad Training the HGNN neural network with ${L}_b$.
\STATE \qquad $r_b \leftarrow r_n$
\STATE \quad \textbf{Else}
\STATE \qquad $\eta \leftarrow \eta \times \tau $	
\STATE \quad \textbf{End If}
\STATE \textbf{End For}
\end{algorithmic}
\end{algorithm}

In machine learning, the generalization ability is a conception that a model can properly adapt to new, previously unseen data that draw from the same distribution as the one used to train the model\cite{bishop_pattern_2006}. In our problem, we define the generalization ability as the ability of an algorithm that can solve the problem for any scenario in the same circumstance. The traditional optimization-based method needs to resolve the optimization problem when the scenario changes. In our method, we define the generalization ability as the neural network can suit for different event of same circumstance. And we fit the neural network with the plenty of training samples of different event to let our network to embrace the generalization ability. However, the representation with good generalization ability could be hard to reach optimal performance in a specific event. The representation should perfectly match this scenario to achieve higher performance. To balance good generalization ability and the performance for particular scenarios, the learning phase of our proposed method is separated into two parts, Generalization-Representation Learning (GRL) part and Specialization-Representation Learning (SRL) part. 

In GRL part, we train the neural network with different training samples to learn the generalization representation. This part of our method for joint user association and power control in HUDNs with semi-supervised learning is summarized in algorithm 2. To eliminate the variance of loss, we update the neural network parameter with the batch sample, which is following the  steps 6 to 12 in algorithm 2. Besides, we dynamically adjust the learning rate according to the variation tendency of rate, as in step 16 to step 20 of Algorithm 2.


The SRL part is aiming to let the neural network perfectly match the particular scenarios. But, applying the representation learning directly in a particular scenario could be inefficient. On the one hand, the loss of a single sample would be undulated, which leads to the algorithm cannot be converged. On the other hand, the number of stagnation points for a single sample is much more than the number of stagnation point for the multisample. To avoid the neural network falling into the local optimum, the parameters of the neural network are finely turned by retraining it with a particular scenario radio map, after the GRL. To further eliminate the local-optimization effect that the overfitting brings and increase exploration, we add the entropy of outputs of the softmax function with temperature parameter $T=1$ to keep the outcome from too confident\cite{sohn_fixmatch_2020}, which is

\begin{equation}
\begin{aligned}
e_{n k} &=\frac{\exp \left(z_{n k}\right)}{\sum_{i \in \mathcal{N}} \exp \left(z_{j k}\right)} \\
{L}_{e n t r o p y} &=-\sum_{j \in \mathcal{K}_{s}} \sum_{i \in \mathcal{N}} e_{i j} \ln e_{i j}.
\end{aligned}
\end{equation}

Besides, we use smaller $w_s$ to let the neural network to explore more effective associate way. The SRL part is summarized in algorithm 3. The differences between SRL and GRL is that the SRL is trained by single radio map and update the parameter with single sample.  
 
Then, the loss function for SRL is:
\begin{equation}
{L}= w_s{L}_{s}-w_rR-w_e{L}_{e n t r o p y},
\end{equation}
where $w_e$ is weight for the loss of entropy.

\begin{algorithm}
\caption{Generalization Representation Learning Algorithm}
\label{alg2}
\begin{algorithmic}[1]
\STATE \textbf{Input}: The radio map of particular scenario $\mathcal{M}_p$; learning rate $\eta$; convergence bound $e$
\STATE Load the HGNN neural network and the parameter of HGNN.
\STATE \textbf{while} \textbf{True}:
\STATE \quad ${\bf{h}}_v^0 \leftarrow \mathcal{M}_p$.
\STATE \quad Calculate the user association $\mathbf{x}$ and power control $\mathbf{p}$ configuration thought HGNN \\
\quad neural network.
\STATE \quad Calculate the loss function ${L}$ of (15), and training with
\STATE \quad Add the rate $R$ into set $\mathcal{M}$.
\STATE \quad \textbf{End For}
\STATE \quad Calculate the average rate $r_n$ of recent $l$ elements in memory set $M$.
\STATE \quad  \textbf{If} $|r_n -r_b|<e$
\STATE \qquad \textbf{End while}
\STATE \quad  \textbf{End If}
\end{algorithmic}
\end{algorithm}

Although the general representation can suit for any UE circumstance, the learning process is time-costly. Fortunately, the GRL part is an off-line training process which requires to be retrained only when the radio map updates. Comparatively, the SRL part suits the instantaneous radio map, which is online training. However, because the SRL is trained based on the GRL, the convergence time is much less than that of GRL. Combining the two parts, the strategy is flexible in real implementation. When the scenario changes quickly, the HUDNs can directly apply the outcome of general representation learning without the SRL part. If the HUDNs wants better performance, the retraining part can lead the neural network to perfectly match the scenario. 

\section{NUMERICAL RESULTS}

In this section, we present some simulations to demonstrate the efficiency of the proposed algorithms. As shown in Fig.5, we consider a squared area with a side length of $L$ in an urban environment, wherein $N$ small BSs are randomly distributed in $M$ Macro BSs covering the area. To simulate the city environment, we randomly simulate 20 buildings with 20m lengths, 20m width, and 30m high, which is showed as blue squares in Fig.5. The BSs who locate in a blue square are deployed on the roof of the building. The blue five-pointed star and the red triangle represent the Marco BS and the small BS, respectively. In practice, UEs could be anywhere in the scenario. To simplify the calculation, in our model, we evenly divide the scenario into the grid and assume the UEs are located in the grid. In every adjust time frame, we assume $K_a$ UEs are activated following a mutually independent fixed stochastic probability. Unless otherwise specified, we set $N=100$, $M=5$, $L=200$, $K_a = 120$. 
\begin{figure}[h]
\centering
\includegraphics[width=3.5in]{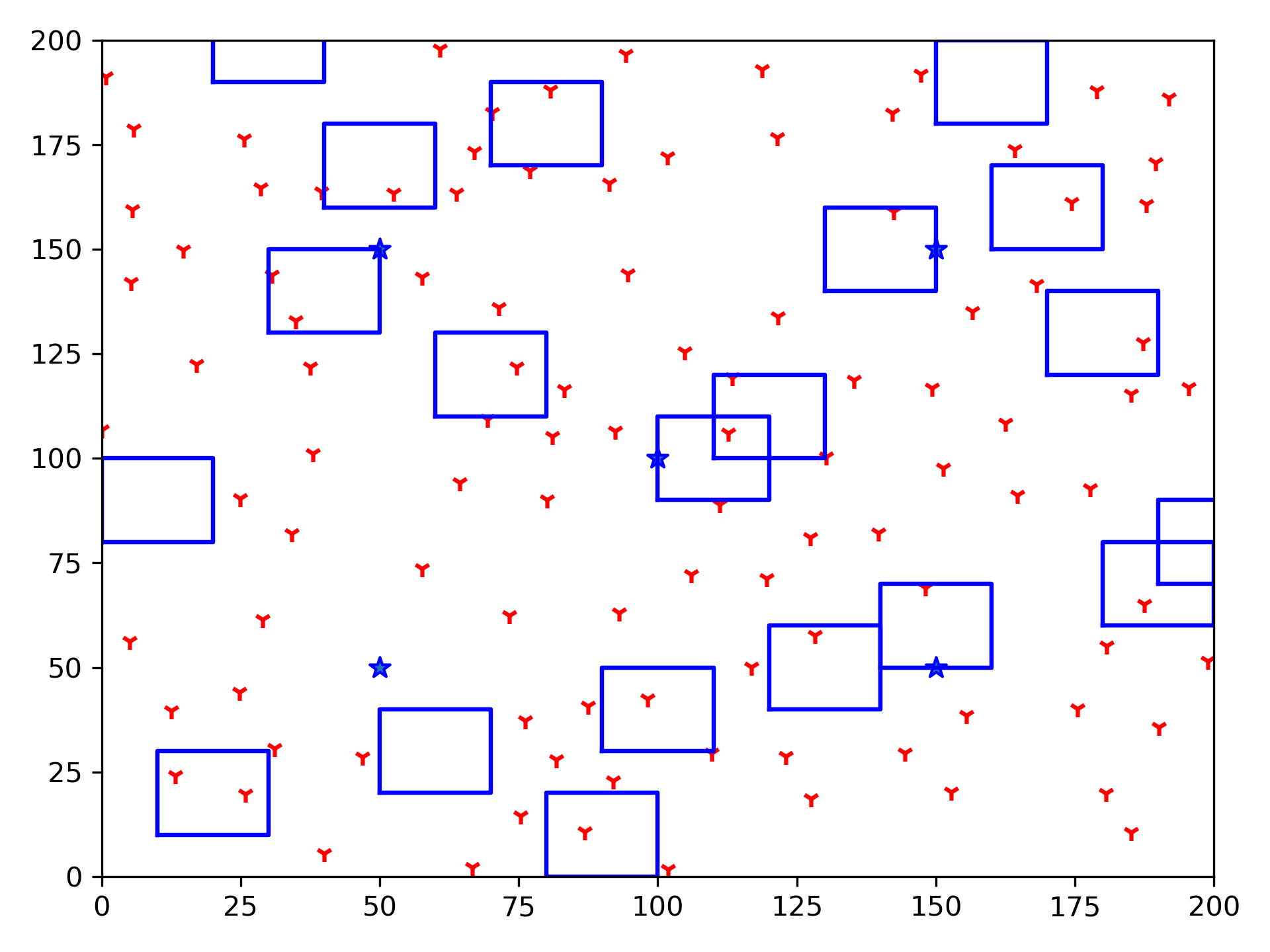}
\caption{The HGNN Architecture of Representation Functions.}
\end{figure}

\begin{center}
\begin{table}
\caption{SIMULATION PARAMETERS}
  \centering
  \begin{tabular}{p{5cm}l}
  \hline
  Parameter & Value/Status \\
  \hline
  Cell area & $200$m $\times$ $200$km\\
  Bandwidth & $20$MHz\\
  Macro BS & 5 \\
  Small BS & 100\\
  The power of Macro BS & 50dBm\\
  The power of small BS & 20dBm\\
  Batch size & 64\\
  Learning Rate & 1e-4\\
  Decay Factor & 0.99\\
  Window length & 100 \\
  Weight of supervised learning & 100 \\
  Weight of unsupervised learning & 1 \\
  Weight of entropy & 1e-2 \\
  $D^{\mathrm{L}}$,$D^{\mathrm{NL}}$ & 10.38,14.54 \\
  $\theta^{\mathrm{L}}$,$\theta^{\mathrm{NL}}$ & 2.09,3.75 \\
  \hline
  \end{tabular}
\end{table}
\end{center}

The channel model that we used is grouped into line-of-sight (LoS) and non-line-of-sight (NLoS) transmission. Because our objective is considered in a long term, we only consider the large-scale channel gain. Then the channel between BS and UE can be obtained by the following equation


\begin{equation}
\zeta(w)=\left\{\begin{array}{ll}
D^{\mathrm{L}} w^{-\theta^{\mathrm{L}}}, & \text { LoS  } \\
D^{\mathrm{NL}} w^{-\theta^{\mathrm{NL}}}, & \text { NLoS} 
\end{array}\right..
\end{equation}
Where $D^{\mathrm{L}}$ and $D^{\mathrm{NL}}$ represent the LoS and NLoS path loss at the unit reference distance, respectively. $\theta^{\mathrm{L}}$ and $\theta^{\mathrm{NL}}$ represent LoS and NLoS path loss indexes, respectively. 

As mentioned in section IV.B, the neural network in our simulation composes two parts, the feature extractor and the output layer. The feature extractor is formed by two HGNN layers. Both layers using 128 neurons for each node. The output layer for power control uses 1 neuron for each node, as for user association, the output layer uses $K$ neutrons to represent the connected index for every BS. The other simulation parameters are listed in Table I.

We compare our proposed method with several schedules, which is Maximal Achievable Rate Association with Maximum Power (MARAMP), Maximal Sum-Utility association with Maximum Power(MSUAMP), that proposed in paper \cite{ye_user_2012}, Maximal Sum-Utility association with Power Control (MSUAPC), User Association with Maximizing the Weighted Sum of Effective Rates(UAMWSER) and the Joint User Association and Power Control with Maximizing the Weighted Sum of Effective Rates (JUAPCMWSER) in paper \cite{zhou_joint_2016}.

%

\subsection{The Convergence Performance of Proposed Method}
We first show the convergence of our proposed method in Fig.6, which concludes the rate, the supervised loss, and total loss for GRL (the Fig6.(a),(c),(e)) and for SRL (the Fig6.(b),(d),(f)), respectively. The Fig.6(a) contain 4 different rates in each training step of GRL, which is the average rate of each batch, the max rate of each batch, the rate combining our PC strategy and MSUA, the rate combining our user association strategy with max transmit power. From the pic, we can see the rate rapidly converge in around 1000 steps and the mean rate of GRL reaches the rate of combination strategy of our power control strategy and MSUAMP. Combine the loss of supervised learning as shown in Fig.6(b), we can get the conclusion that in the first 1000 steps the neural network is mainly trained by supervised learning. The same conclusion can be get from the rate of combination strategy of our user association strategy and max transmit power strategy. After several steps, the neural network is mainly trained by the unsupervised loss. 

\begin{figure*}[!t]
\centering
\subfloat[Rate of GRL part]{\includegraphics[width=2.8in]{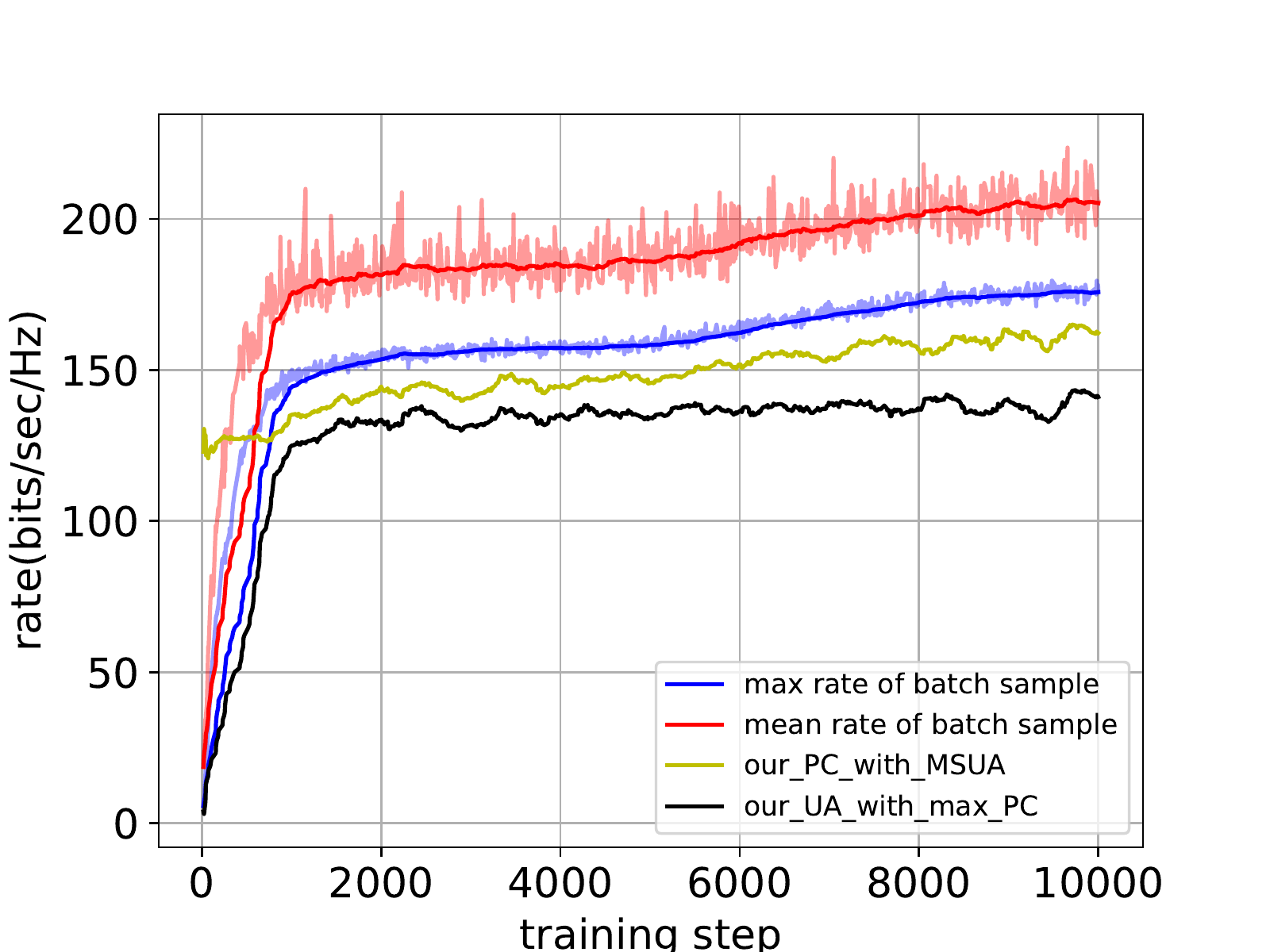}}
\hfill
\subfloat[Rate of SRL part]{\includegraphics[width=2.8in]{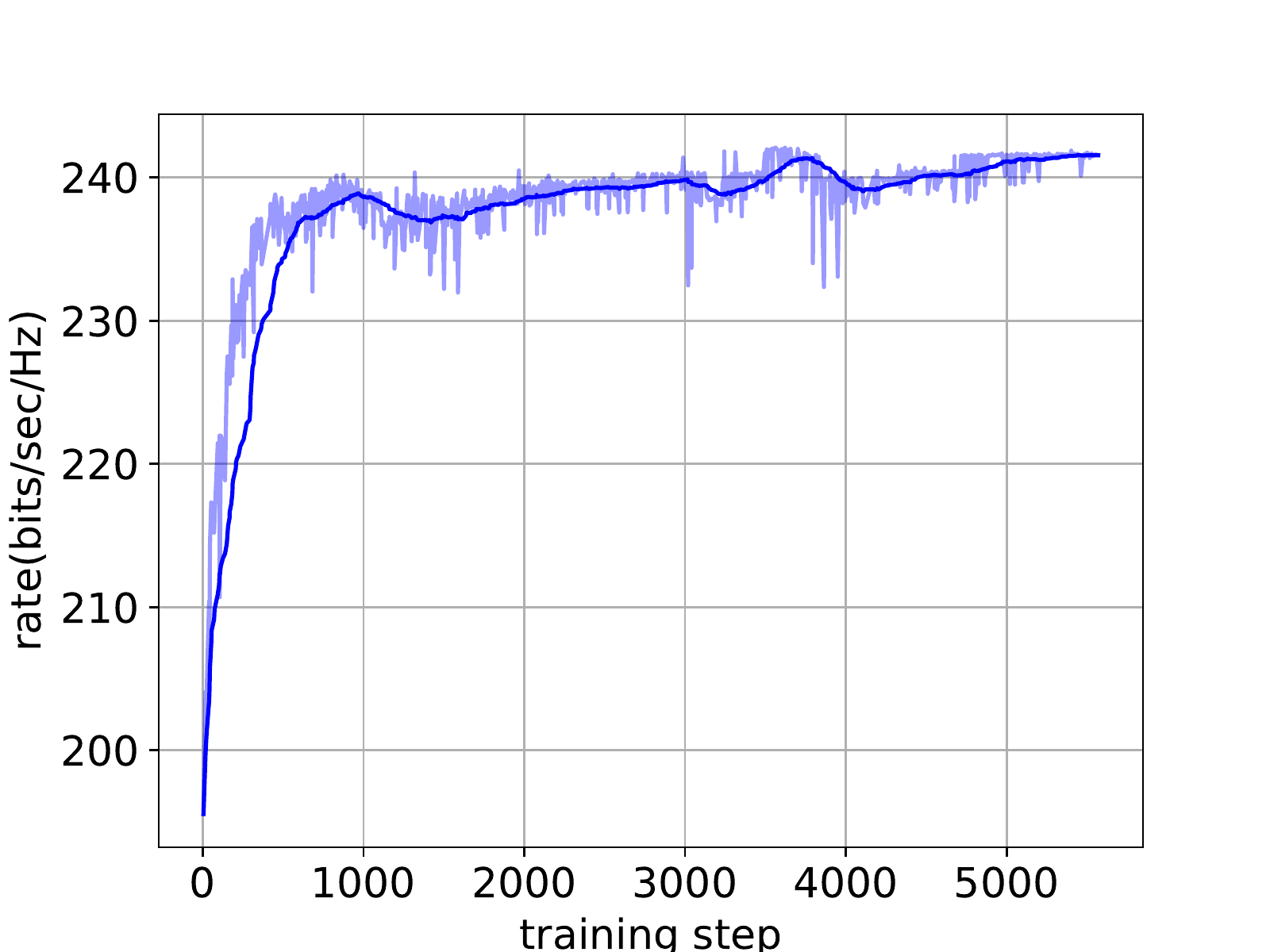}}\\

\subfloat[Supervised loss of general representation learning part]{\includegraphics[width=2.8in]{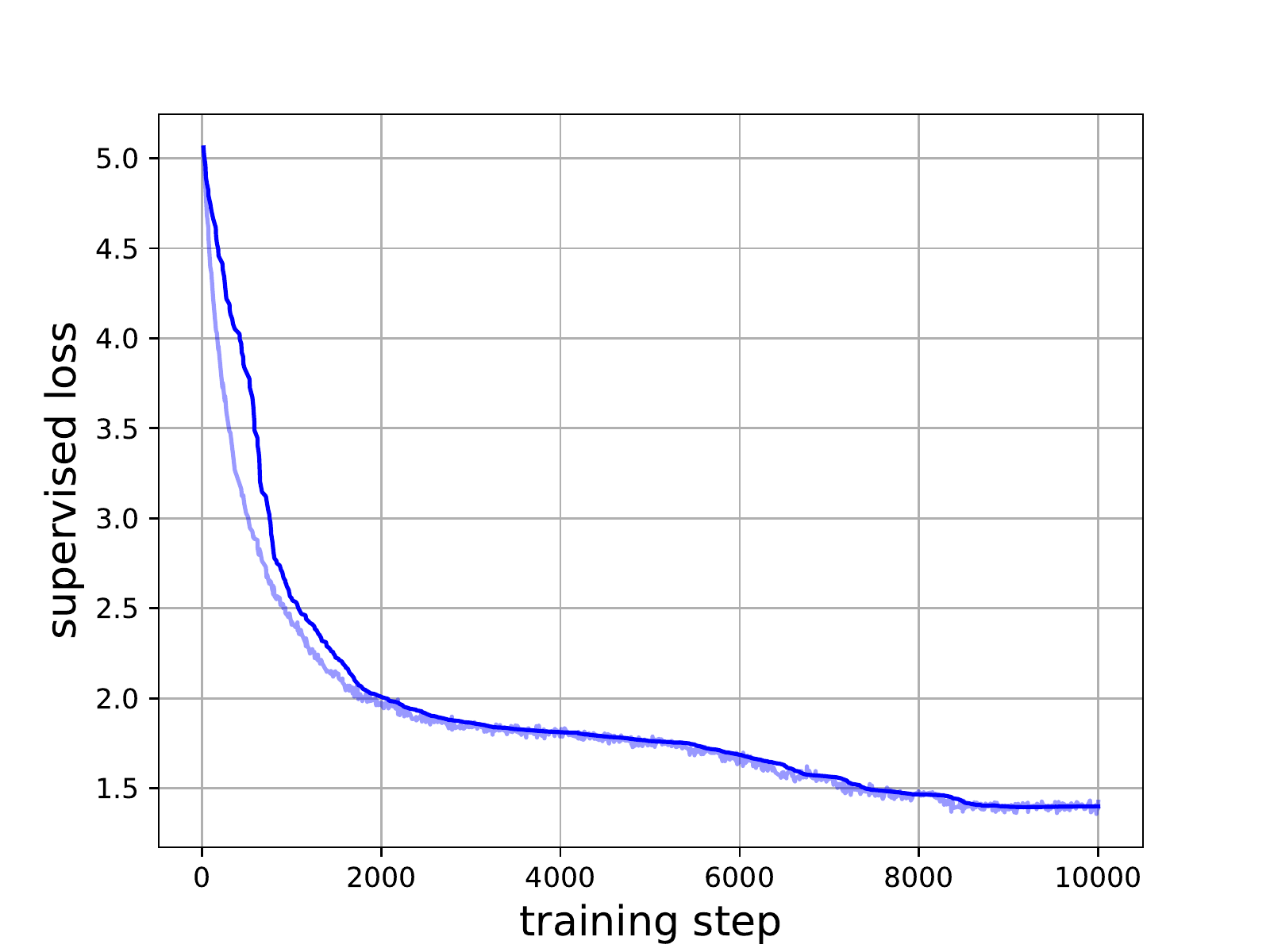}} 
\hfill
\subfloat[Supervised loss of retraining part]{\includegraphics[width=2.8in]{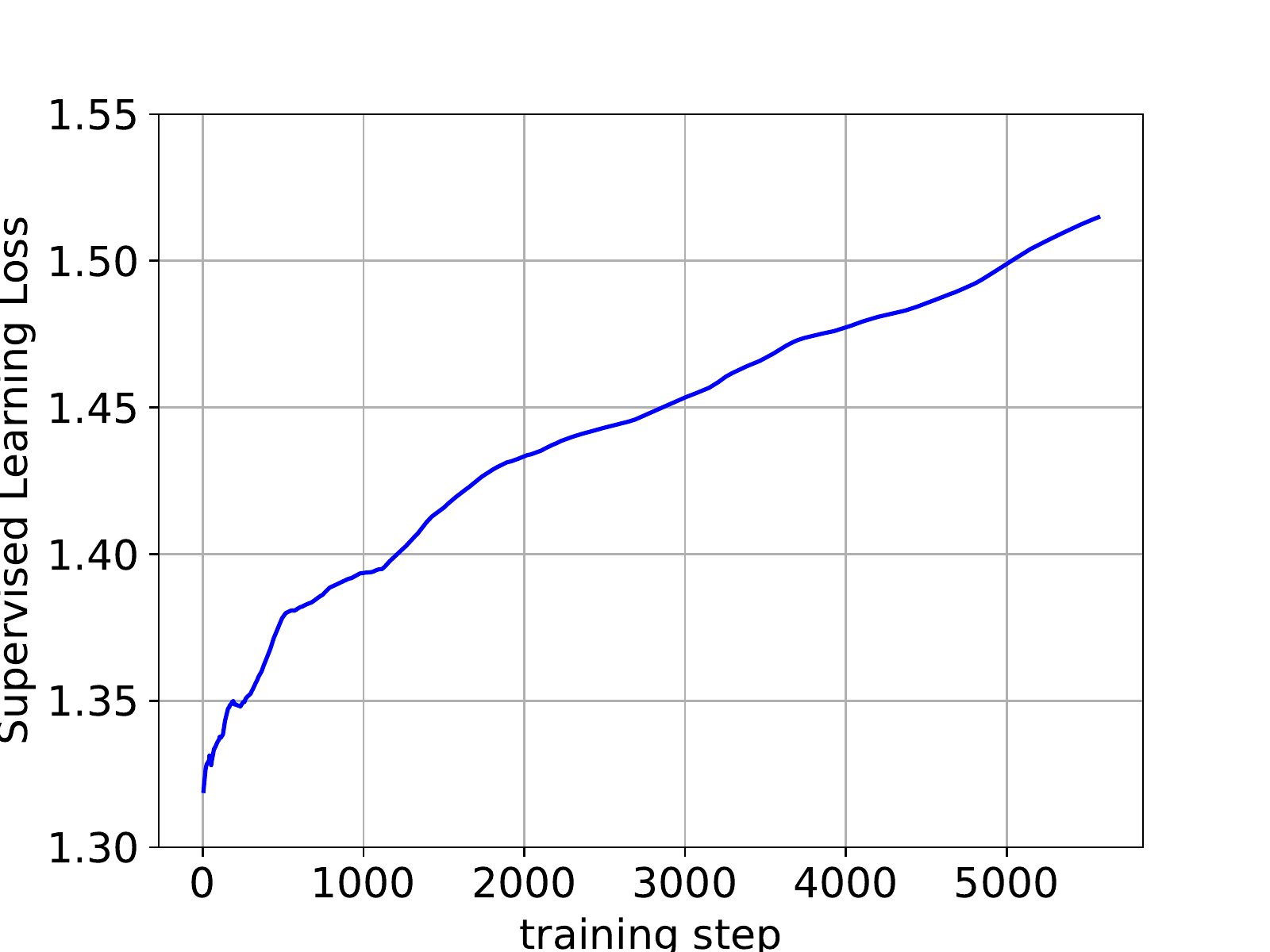}}\\
\subfloat[Total loss of general representation learning part]{\includegraphics[width=2.8in]{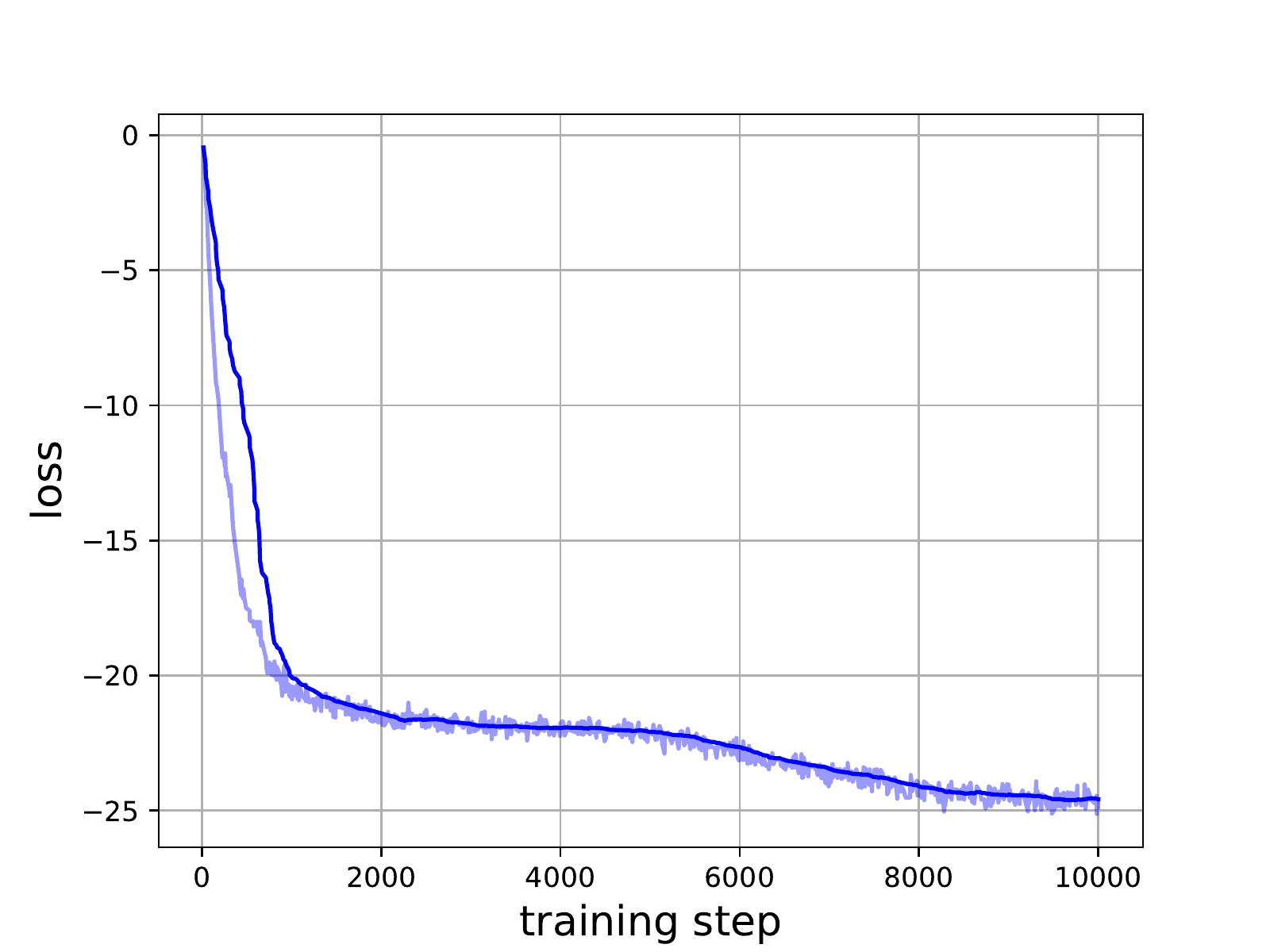}}
\hfill
\subfloat[Total loss of retraining part]{\includegraphics[width=2.8in]{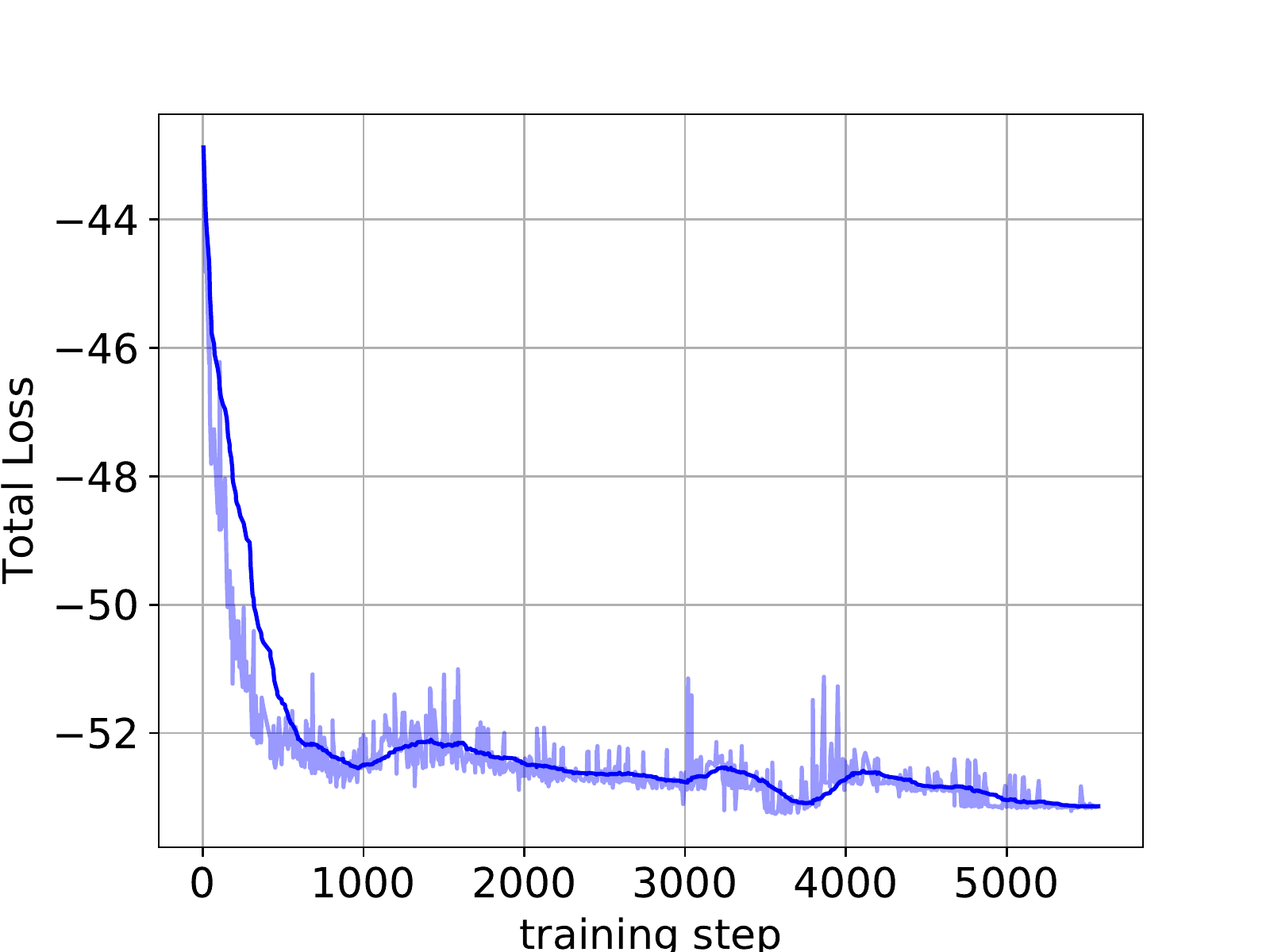}} \\
\caption{The Convergence of proposed method}
\label{fig_sim}
\end{figure*}

The Fig6.(b) shows the SRL part can efficiently improve the performance of GRL in specific scenario through lightly adjusting the user association and power control, which is also reflected in the Fig6.(d) and the Fig6.(f). Due to the small weight of supervised loss in the SRL part, the supervised learning loss is raised because the unsupervised learning loss plays a leading role and lead the neural network to find more appropriate cooperation of user association and power control strategy instead of constrained by the given label.  

\subsection{The Performance Analysis}

In this section, we show the performance of the mentioned algorithm from the total effective rate, the cumulative distribution function (CDF) of rate, which is shown in Fig.7, Fig. 8, respectively. Also, we compare our algorithm in different UE density. We simulate 200 times for each algorithm.

\begin{figure}[h]
\centering
\includegraphics[width=4.5in]{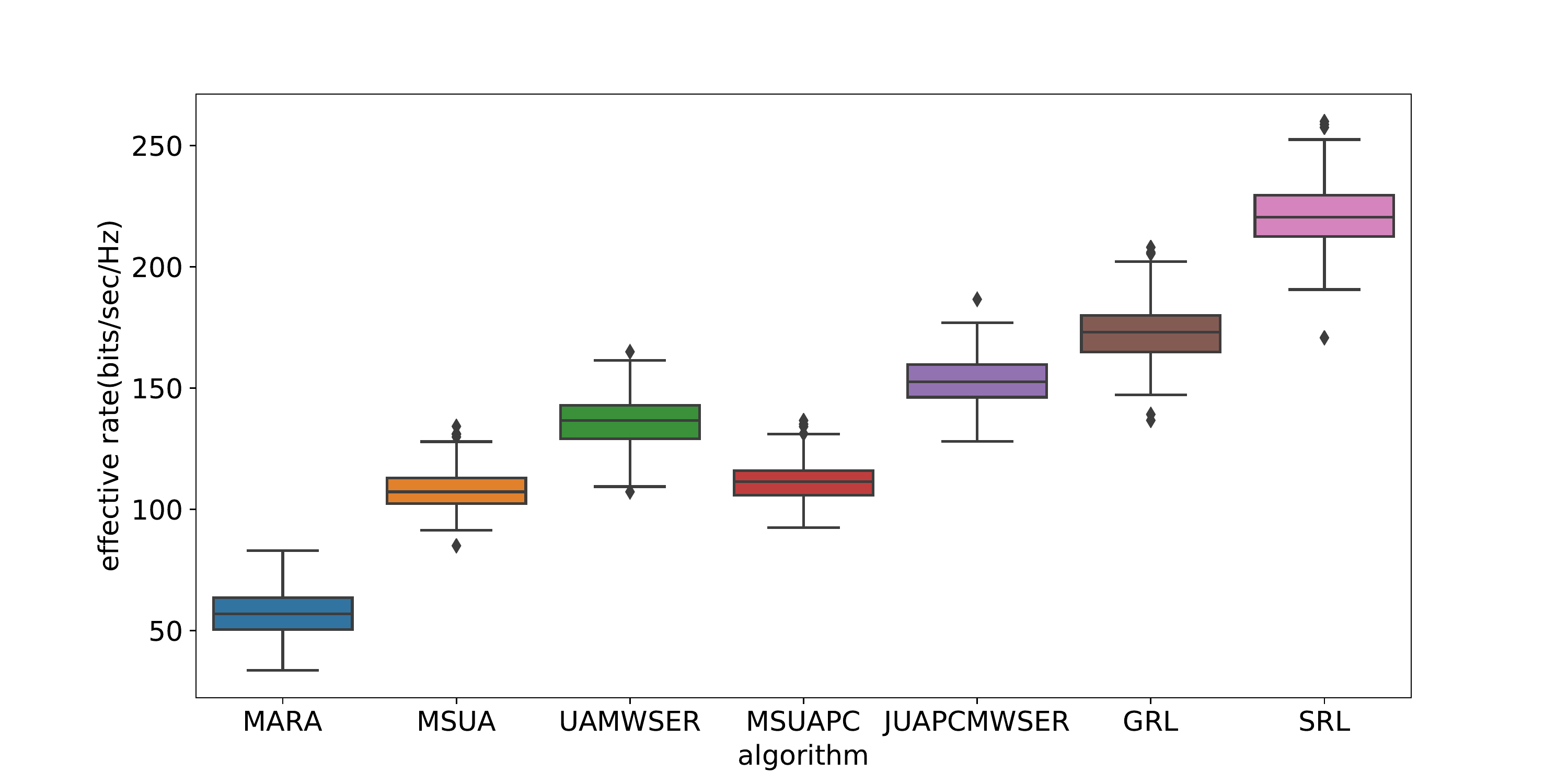}
\caption{The average rates of different association strategies..}
\end{figure}

\begin{center}
\begin{table}
\caption{Total rate of each algorithms}
  \centering
  \begin{tabular}{p{3cm}p{1.5cm}p{1.5cm}p{1.5cm}p{1.5cm}p{2cm}p{1.5cm}c}
  \hline
  Algorithm & MARAMP & MSUAMP & MSUAPC & UAMWSER & JUAPCMWSER & \bf{GRL} & \bf{SRL} \\
  \hline
  Total rate(bit/sec/Hz) & 57.259 & 108.081& 136.219 & 111.266 & 153.227 & \bf{172.565} & \bf{221.000} \\
  \hline
  \end{tabular}
\end{table}
\end{center}

As illustrated in Fig.7, the MARAMP has the most low-rate users among all strategies since all of UE connect to the Macro BS according to the signal strength. The MSUAMP and UAMWSER show better performance because the connection is more balancing, but the UE in this scenario suffers from strong interference from Macro BS. With active power configuration structure, the MSUAPC and JUAPCMWSER perform batter than MSUAMP and UAMWSER, which reach 136.219 bit/sec/Hz and 153.227 bit/sec/Hz respectively. Our algorithms show an overwhelming superiority over other algorithms. The GRL and SRL achieve 172.565 bit/sec/Hz and 221.000 bit/sec/Hz. 

In fig.8, the same conclusion can be get. The MARAMP have the UE that reach the highest rate but have the most the lower-rate UE as well. The performance of MSUAMP, UAMWSER, MSUAPC is similar. Compared with them, the JUAPCMWSER shows better balance and achieves a higher max effective rate. The GRL is a prime performance in load balancing. The GRL and SRL show different characters. The SRL sacrifices some UEs' rate but to improve the total effective rate, however, the result of GRL is more equal for every UE, which proofs the UA label that is used by the supervised learning is relatively equal UA strategy. The unsupervised learning part prefers to achieve a higher total rate but partly ignore the load balance.

\begin{figure}[h]
\centering
\includegraphics[width=3.5in]{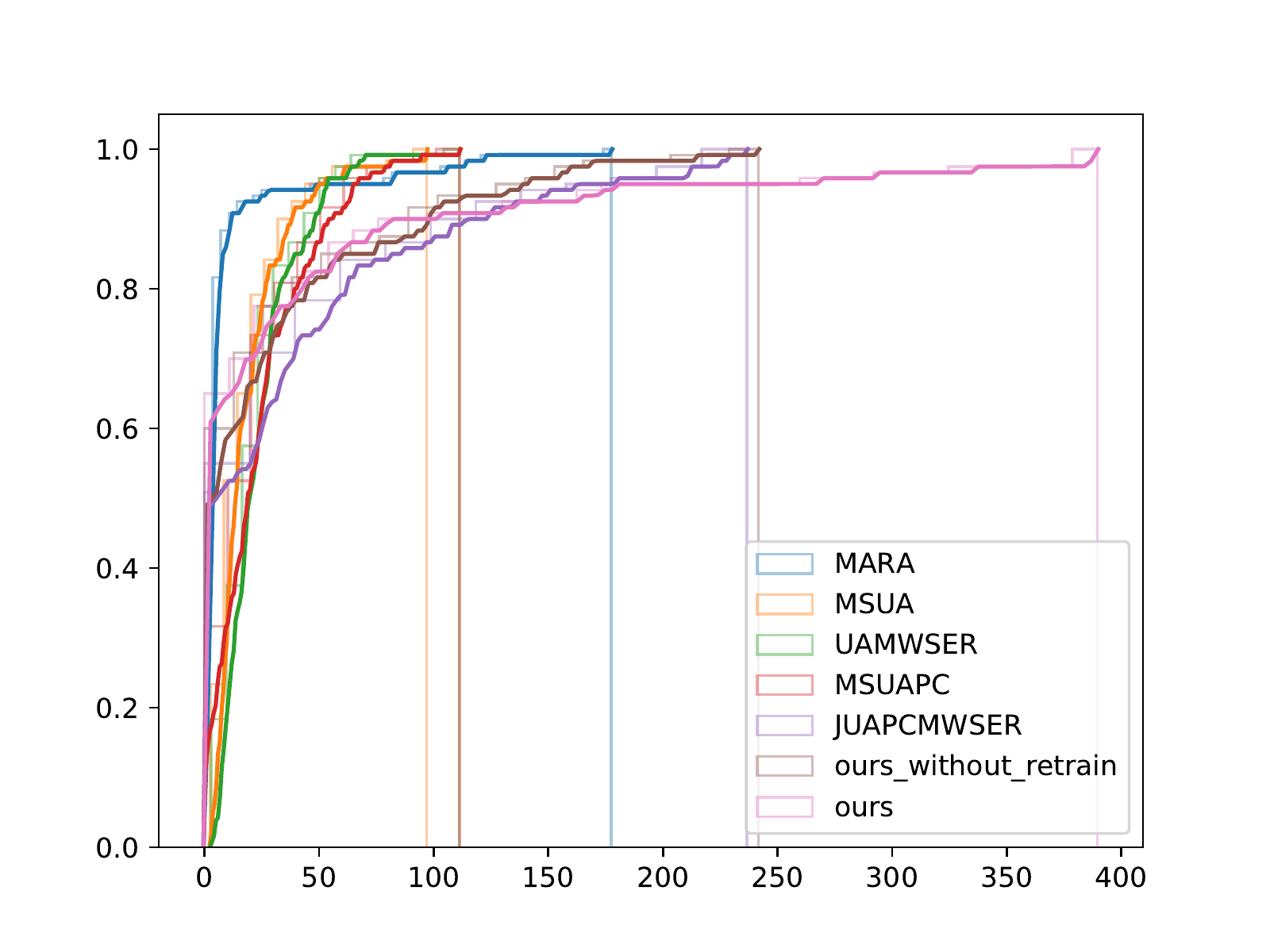}
\caption{The CDFs of effective rates of associated users under different association strategies.}
\end{figure}

\begin{figure*}[!t]
\centering
\subfloat[GRL]{\includegraphics[width=3.0in]{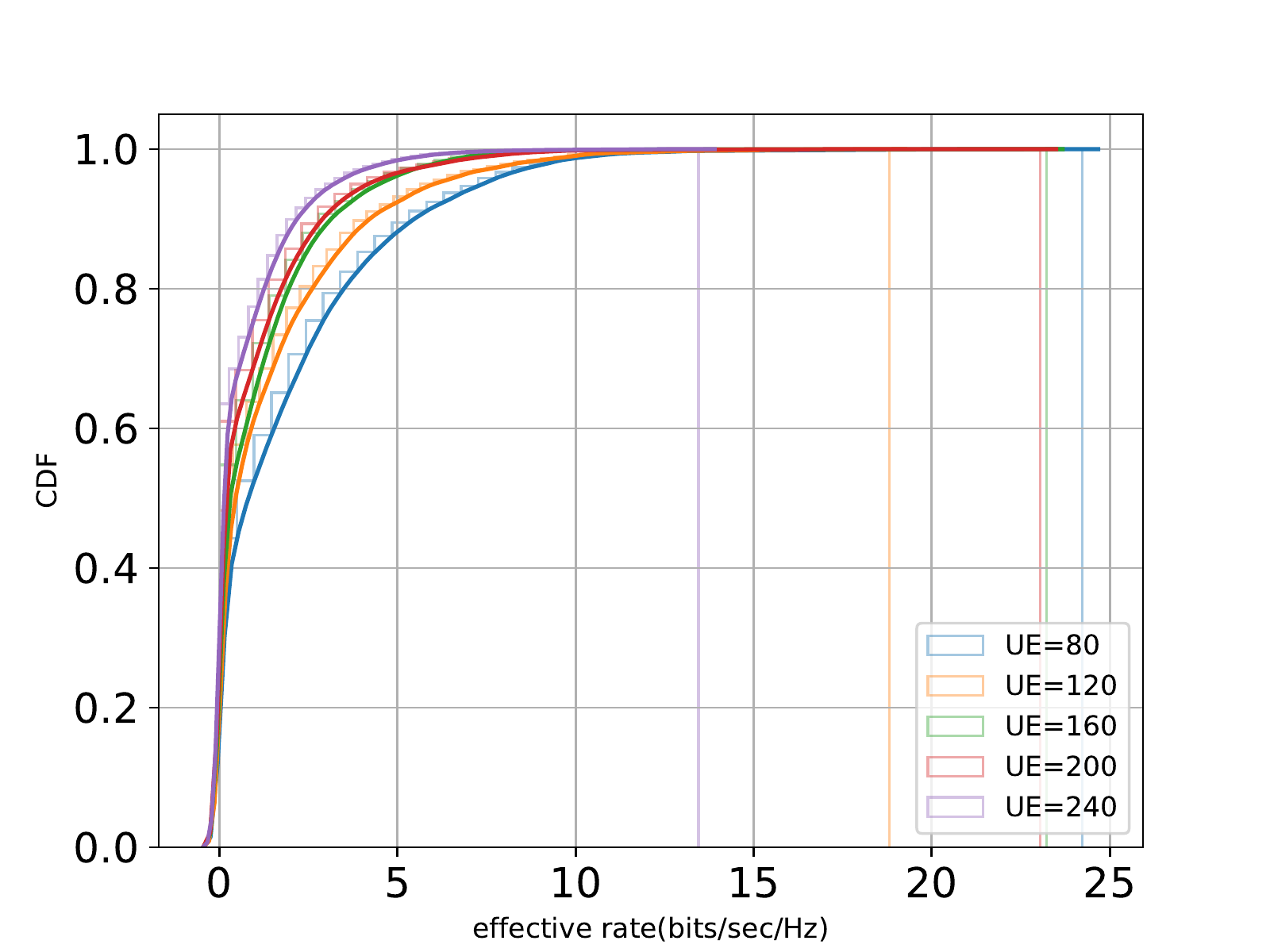}}
\hfill
\subfloat[SRL]{\includegraphics[width=3.0in]{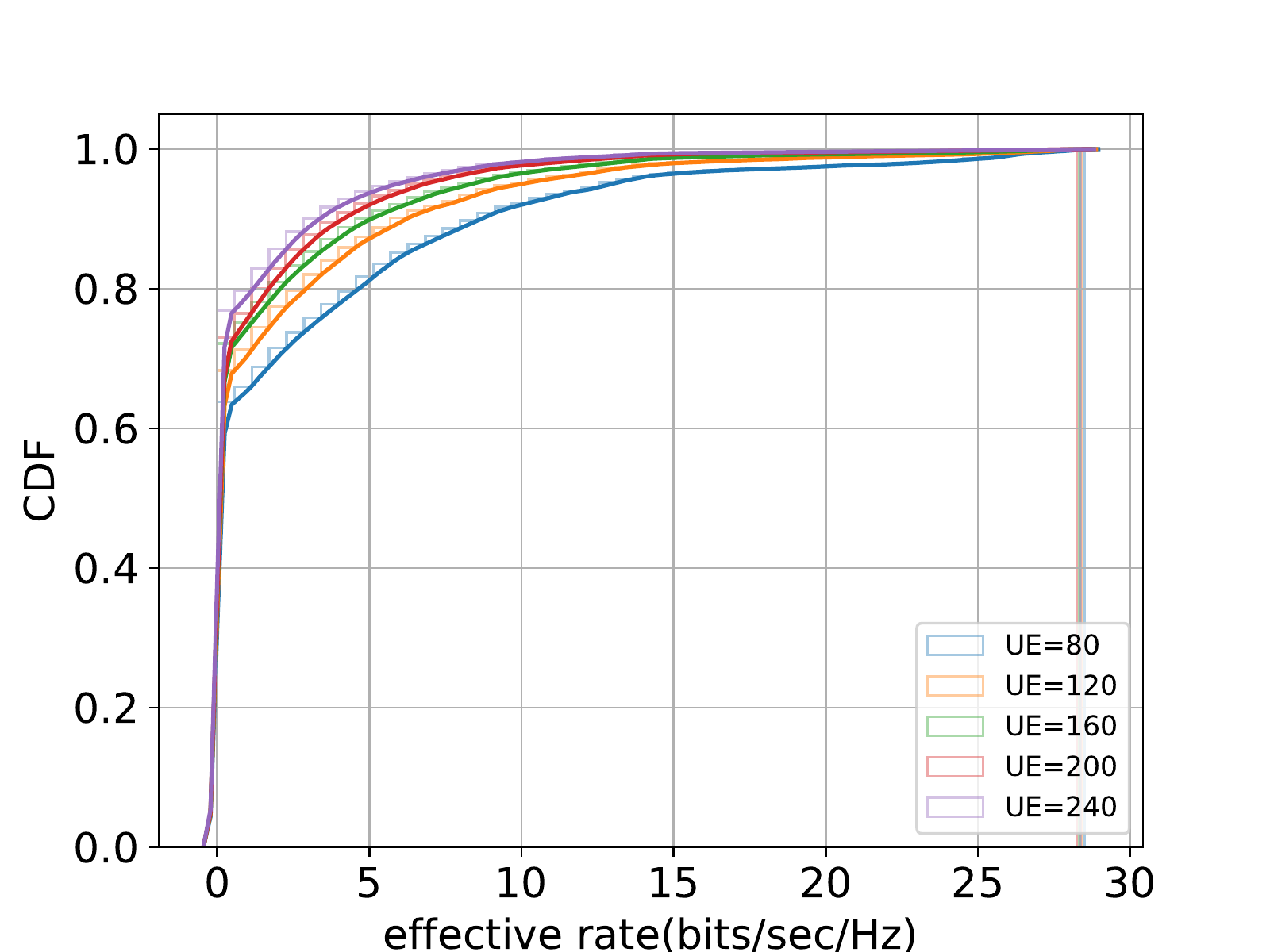}}\\
\caption{The performance of our method in different UE density}
\label{fig_sim}
\end{figure*}

The Fig.9 show the CDF in different UE density, which prove our algorithms show better performance in low UE density for less interference. Besides, the same conclusion that the result of SRL preference to maintain the UE that can reach the higher rate to the boost overall performance. It is easy to conclude that the max effective rate is almost same in the SRL.

\subsection{The Comparison of Computational Efficiency}

To compare the computational efficiency of all 7 algorithms, we count the calculation time of each algorithm. We test in 200 events with python on the workstation with Intel Core i7-9700K. 

The result is concluded in table II. A clear conclusion can be got that the algorithm with the iteration process consumes more computation time. The MARAMP and GRL show excellent performance that allows the communication system to handle in a short moment. The GRL only needs several matrix multiplication operations when solving the problem after the learning part. Compared to the GRL, the optimization-base problem need more than 10 times of computation to converge. Especially for the joint optimization problem, the multilayer iteration process of two optimal problems exponentially increases the computation time. As for SRL, the learning process can be also treated as an iteration process. It also needs time to converge. But the SRL only has one layer iteration, most of the retaining time is used in back-propagation and the adjusting of network's parameter.

\begin{center}
\begin{table}
\caption{Computational Efficiency}
  \centering
  \begin{tabular}{p{2cm}p{1.5cm}p{1.5cm}p{1.5cm}p{1.5cm}p{2cm}p{1.5cm}c}
  \hline
  Algorithm & MARA & MSUA & MSUAPC & UAMWSER & JUAPCMWSER & \bf{GRL} & \bf{SRL} \\
  \hline
  Mean time(s) & 0.037 & 3.334 & 285.503 & 0.204 & 1019.489 & \bf{0.039} & \bf{97.697} \\
  \hline
  \end{tabular}
\end{table}
\end{center}

\section{Conclusion}
A joint user association and power control framework was proposed for HUDNss aiming at maximizing the total effective rate. We built the HUDNs as a heterogeneous graph and trained the Graph Neural Network with semi-supervised learning to learn a generalized representation as a resource allocation scheme. Through decomposing the training processing into two parts, our proposed method was able to suit for the rate-required situation and speed-require situation. Our proposed user association and power control solution offered the superior performance of load balancing, total effective rate, and calculation efficiency under a range of practical system settings. For future work, we will consider designing the loss function that can train the GNN more efficiently. Besides, instead of a total effectivedate rate, different performance indicators or multi-component objective functions will be considered in HUNDs. Also, different loss functions will be considered to be used in the GRL part and SRL part to pursuit different performance indicators.



%

%
%

\ifCLASSOPTIONcaptionsoff
  \newpage
\fi



\bibliographystyle{IEEEtran}
\bibliography{2020,other11.bib,other.bib}
\end{document}